\documentclass[useAMS,usenatbib,usegraphicx]{mn2e}
\usepackage{psfig}   
\usepackage{graphicx}
\usepackage{amssymb}
\usepackage{subfigure}

\title[Characterizing the state of clusters]{Characterizing the dynamical state of star clusters from snapshots of their spatial distributions}

\author[R.~J.~Parker \& M.~R.~Meyer ]{
  Richard J.~Parker\thanks{E-mail: rparker@phys.ethz.ch} and Michael R. Meyer \vspace*{0.1cm}\\
   Institute for Astronomy, ETH Z{\"u}rich, Wolfgang-Pauli-Strasse 27, 8093, Z{\"u}rich, Switzerland}

\begin{document}

\date{}
                             
\pagerange{\pageref{firstpage}--\pageref{lastpage}} \pubyear{2012}

\maketitle

\label{firstpage}

\begin{abstract}
We determine the distribution of stellar surface densities, $\Sigma$, from models of static and dynamically evolving star clusters with different morphologies, including both radially smooth and 
substructured clusters. We find that the $\Sigma$ distribution is degenerate, in the sense that many different cluster morphologies (smooth or substructured) produce similar 
cumulative distributions. However, when used in tandem with a measure of structure, such as the $\mathcal{Q}$-parameter, the current spatial and dynamical state of a star cluster can be inferred. 
The effect of cluster dynamics on the $\Sigma$ distribution and the $\mathcal{Q}$-parameter is investigated using $N$-body simulations and we find that, depending 
on the assumed initial conditions, the $\Sigma$ distribution can rapidly evolve from high to low densities in less than 5\,Myr. This suggests that the $\Sigma$ 
distribution can only be used to assess the current density of a star forming region, and provides little information on its initial density. However, if the $\Sigma$ distribution is used together with the 
$\mathcal{Q}$--parameter, then information on the amount of substructure can be used as a proxy to infer the amount of dynamical evolution that has taken place. Substructure is erased quickly through dynamics, 
which can disrupt binary star systems and planets, as well as facilitate dynamical mass segregation. Therefore, dynamical processing in young star forming regions could 
still be significant, even without currently observed high densities.
\end{abstract}


\section{Introduction}

Stars form in clusters and associations  \citep[e.g.][]{Carpenter00b,Lada03,Lada10}, and their spatial distribution appears to be smooth and continuous, i.e.\,\,there is no evidence of bi--modal star formation \citep{Bressert10}. 
However, the morphologies of individual clusters appear to vary significantly \citep{Cartwright04}, with some regions displaying a high amount of substructure (e.g.~Taurus and Chamaeleon) and others displaying 
a smooth, centrally concentrated morphology (e.g.~$\rho$~Ophiuchus and IC\,348). This suggests that either star formation in clusters is not universal and may depend on the local environment, or that all clusters 
form with the same morphology which is later altered by dynamical interactions.

There are several methods to look for structure in clusters, and even clusters themselves. The study by \citet{Cartwright04} used a minimum spanning tree analysis, which has also been used to quantify the amount of mass segregation in clusters 
\citep{Allison09a} and can be used to search for clusters, or ``clustering'' in crowded fields \citep{Gutermuth09,Schmeja11}, in tandem with the mean separation between stars, to define the $\mathcal{Q}$--parameter. The $\mathcal{Q}$--parameter 
quantifies whether the cluster is substructured or radially concentrated, and to what extent, for a given morphology.

The stellar surface density distribution has also been used to study the structure of star forming regions. \citet{Larson95} showed that the structure in Taurus was hierarchical, but broke down at the binary 
regime (corresponding to the local Jeans length), and \citet{Simon97} found the same result for Ophiuchus and the Trapezium cluster. However, \citet*{Bate98b} repeated this analysis for the Trapezium cluster and found that the 
break between random stars and binaries does not correspond to the Jeans length in this cluster.  Furthermore, \citet{Kraus08} found that the break between random stars and binaries was 11\,000\,au in Upper Sco, compared to 17\,000\,au in Taurus, 
and suggested the break is a function of the age (and initial density) of the cluster, and is caused by dynamical evolution.

Recently, the distribution of stellar surface densities  ($\Sigma$) was also used to evaluate the definition of what constitutes a star cluster. \citet{Bressert10} examined a volume-limited sample of nearby star-forming regions and calculated the distribution 
of $\Sigma$ in each region. \citet{Bressert10} then compared the cumulative distribution of $\Sigma$ for all the regions (apart from the central region of the Orion Nebular Cluster (ONC), where the determination of 
$\Sigma$ was not possible for every star due to crowding) to various definitions of clustering, based on stellar surface density thresholds. Due to a lack of obvious bi-modality in the distribution, 
they concluded that stars in the local solar neighbourhood form in a continuous spatial distribution. 

In this paper, we determine the effectiveness of the surface density distribution, $\Sigma$, in tandem with the $\mathcal{Q}$-parameter, for providing information on the dynamical state of star clusters. We begin by analysing static star clusters with various morphologies to look for variations in the shape of the $\Sigma$ distribution.  
We then dynamically evolve radially smooth 
Plummer sphere clusters, and substructured fractal clusters, and determine the $\Sigma$ distribution and the $\mathcal{Q}$-parameter as a function of time. We describe the set-up of the clusters in Section~\ref{method}; we present our results in Section~\ref{results}; we 
provide a discussion in Section~\ref{discuss} and we summarize and conclude in Section~\ref{conclude}.

\section{Method}
\label{method}

In this section we describe the measurement of stellar surface density, $\Sigma$, and the $\mathcal{Q}$-parameter, which is used to quantify the amount of substructure in star cluster. We then describe the Monte Carlo set-up of static star clusters and associations 
before describing the set-up of $N$-body simulations in which we measure $\Sigma$ and the $\mathcal{Q}$-parameter in star clusters as a function of time.  

\subsection{The stellar surface density distribution, $\Sigma$}

The recent work by \citet{Bressert10} measured the stellar surface density, $\Sigma$, of companions in a volume-limited sample of star forming regions. They adopt the following formula \citep{Casertano85}:
\begin{equation}
\Sigma = \frac{N - 1}{\pi D^2_N},
\end{equation}
where $N$ is the $N^{\rm th}$ nearest neighbour and $D_N$ is the projected distance to that nearest neighbour. The value of $\Sigma$ can be sensitive to the choice of $N$; for example a low $N$ value would cause an artificially high $\Sigma$ value 
if the binary fraction were high. A high $N$ value would fail to detect localised pockets of high density, which can have important implications for the evolution of substructured clusters \citep*{Parker11c}. \citet{Bressert10} adopt $N = 7$, which is also the 
value we choose.

\subsection{The $\mathcal{Q}$-parameter}

The $\mathcal{Q}$-parameter was pioneered by \citet{Cartwright04} and combines the normalised mean edge length of the minimum spanning tree of all the stars in the cluster, $\bar{m}$, with the normalised correlation length between all stars in the cluster, 
$\bar{s}$. The level of substructure is determined by the following equation:
\begin{equation}
\mathcal{Q} = \frac{\bar{m}}{\bar{s}}.
\end{equation}
A substructured cluster has $\mathcal{Q}<0.8$, whereas a smooth, centrally concentrated cluster has $\mathcal{Q}>0.8$. The $\mathcal{Q}$-parameter has the advantage of being independent of the density of the star forming region, and purely measures 
the level of substructure present. The original formulation of the $\mathcal{Q}$-parameter assumes the cluster is spherical, but can be altered to take into account the effects of elongation \citep{Bastian09}.

\subsection{Cluster morphologies}
\label{synthetic}

We set up clusters with four different morphologies. Two are radially smooth, centrally concentrated morphologies -- a Plummer sphere \citep{Plummer11}, and a King profile \citep{King62}. Both are used extensively as the initial conditions of 
$N$-body simulations and assume a relaxed, spherical symmetry. We would therefore expect that these clusters would display a smooth, continuous distribution in stellar 
surface density.

The remaining two morphologies are intended to mimick the observations of young, dynamically unevolved star forming regions (like the majority of those in the \citet{Bressert10} sample), which show a high level of substructure 
\citep[e.g.][and references therein]{Gutermuth09,Sanchez09,Schmeja11}.  Firstly, we adopt the fractal distribution. Secondly we create `associations', with several `nodes' of star formation. 

For each morphology, we create a cluster with $N = 1000$ stars, and then find the physical size scale (i.e.\,\,cluster radius) for that morphology which has the best overlap with the observed distribution in \citet{Bressert10}. To this end, we simply choose the cluster radius so that the median $\Sigma$ best corresponds to the median value in the Bressert et al.\,\,distribution ($\tilde{\Sigma}_{\rm YSO} = 22$\,stars\,pc$^{-2}$). The random number seed used to generate the positions of stars in our model clusters results in some variation around this median, but we are examining the morphologies of these clusters, rather than attempting to produce an exact fit to the observed data. Note also that our chosen sample of stars is smaller than the 
3857 used to make the cumulative distribution in \citet{Bressert10}. The $\Sigma$ distribution in Bressert et al.\,\,is the sum of 12 different star forming regions, 
with a median membership of $N \sim 100$, but is dominated by the Orion complex (2696 objects).  All we aim to investigate here is which, if any, single morphology could feasibly reproduce the observed $\Sigma$ distribution.

\subsubsection{Plummer spheres}

Plummer spheres \citep{Plummer11} were first used to describe the radial profiles of Globular clusters, and are used extensively to model the initial mass distribution in star clusters in $N$-body 
simulations \citep[e.g.][]{Kroupa08}, due to their simplicity. The mass enclosed in radius $r$ is given by the following equation:  
\begin{equation}
M(r) = M_{\rm clus}\frac{\left(\frac{r}{r_{\rm pl}}\right)^3}{\left[1 + \left(\frac{r}{r_{\rm pl}}\right)^2\right]^\frac{3}{2}},
\end{equation}
where $ M_{\rm clus}$ is the total mass of the cluster, and $r_{\rm pl}$ is the characteristic `Plummer radius'. Formally, the Plummer sphere extends to an infinite distance, but the radius within 
which half the mass resides, $r_{1/2}$, is given by $r_{1/2} = 1.305r_{\rm pl}$ \citep{Kroupa08}.

\subsubsection{King profiles}

King profiles \citep{King62} are a more realistic description of of the radial profile of Globular clusters. The King profile is used extensively in $N$-body simulations, but also to fit the profiles of observed clusters, for example the ONC \citep{Hillenbrand98}.  
A core radius, $r_c$ (and central surface density, $f_0$) are defined in the model, and the point at which the radius of the cluster is tidally truncated, $r_t$.  
The surface density, $f$ at a distance $r$ from the centre of the cluster is given by the following relation: 
\begin{equation}
f = \frac{f_0}{1 + (r/r_c)^2}.
\end{equation}
King profiles can be described by a central `concentration parameter', $W_0$ (with $W_0 \propto {\rm log_{10}}(r/r_t)$), and the truncation radius, $r_t$ \citep{King66}. If a King profile has a large value ($> 3$) of $W_0$ then it has an extended envelope, whereas a King profile with a low $W_0$ is said to just have a central core. 
We will detail our choice of $W_0$ and $r_t$ in the following Section.

\subsubsection{Fractals}
\label{fractal}

The fractal distribution has the advantage that the substructure is described by one parameter, the fractal dimension, $D$. In three dimensions, a cluster with fractal dimension $D = 1.6$ is very clumpy, whereas a cluster with $D = 3.0$ is a uniform sphere. 
We set our fractal clusters up according to the method outlined in \citet{Goodwin04a}. This begins by defining a cube of side $N_{\rm div}$ (we adopt $N_{\rm div} = 2.0$ throughout), inside which the fractal is built. A first-generation parent is placed at the 
centre of the cube, which then spawns $N^3_{\rm div}$ subcubes, each containing a first-generation child at its centre. The fractal is then built by determining which of the children themselves become parents, and spawn their own offspring. This is 
determined by the fractal dimension, $D$, where the probability that the child becomes a parent is given by $N^{(D-3)}_{\rm div}$. For a lower fractal dimension, fewer children mature and the final distribution contains more substructure. Any children that do not become parents in a
given step are removed, along with all of their parents. A small amount of noise is then added to the positions of the remaining children, preventing the cluster from having a gridded appearance, and the children then become parents of the next generation. Each 
new parent then spawns $N^3_{\rm div}$ second-generation children in $N^3_{\rm div}$ sub-subcubes, with each second-generation child having a $N^{(D-3)}_{\rm div}$ probability of becoming a second-generation parent. This process is repeated until there are substantially more children than
required. The children are pruned to produce a sphere from the cube  and are then randomly removed (so maintaining the fractal dimension) until the required number of children is left. These children then become stars in the star cluster.

The fractal is therefore described by two numbers, the fractal dimension $D$, and the radius $r_F$. It is not clear 
whether star formation actually gives a fractal distribution in nature \citep[see e.g.][]{Elmegreen01}, but it is the most convenient method for creating substructure. 

\subsubsection{Associations}

We also create a substructured association, by randomly placing 5 `nodes' of star formation in three dimensions. From each node, we distribute stars in a spherical 
coordinate system, defining a characteristic length scale, or `compactness', $C$ which we divide the distance from the centre of each node $N$ by:
\begin{equation}
  r_x =N_x + \frac{XR}{C}{\rm sin}\theta \, {\rm cos}\phi,
\end{equation}
\begin{equation}
 r_y =N_y + \frac{XR}{C}{\rm sin}\theta \, {\rm sin}\phi,
\end{equation}
\begin{equation}
 r_z =N_z + \frac{XR}{C}{\rm cos}\theta,
\end{equation}
where $X$ is a random number between 0 and 1, $\theta \in [0, \pi]$, $\phi \in [0, 2\pi]$ and $R$ is the radius of the association.\\

For each cluster morphology we determine $\Sigma$ for each star and make a cumulative distribution of the $\Sigma$ values. We then determine the $\mathcal{Q}$-parameter for each cluster. Note that we `observe' our clusters in the x--y plane; for each cluster we have checked that the results do not depend on viewing angle (any difference is of order a few per cent -- a similar effect to changing the random number seed used to generate the positions). We detail our choices of cluster 
parameters (e.g.\,\,$r_{1/2}$, $D$, $W_0$, etc.) in Section~\ref{static_results}.

\subsection{Dynamical evolution}
\label{temporal}

We study effects of dynamical evolution on the distribution of $\Sigma$ and the $\mathcal{Q}$-parameter for two types of cluster morphology -- a radially smooth and centrally concentrated Plummer sphere, and a substructured fractal.
Both the Plummer sphere and the fractal comprise of $N = 1000$ single stars with masses drawn from a 2--part power law \citet{Kroupa02} IMF with the following parameters:
\begin{equation}
 N(M)   \propto  \left\{ \begin{array}{ll} 
 M^{-1.3} \hspace{0.4cm} m_0 < M/{\rm M_\odot}  \leq m_3   \,, \\ 
 M^{-2.3} \hspace{0.4cm} m_1 < M/{\rm M_\odot} \leq m_2   \,,
\end{array} \right.
\label{imf}
\end{equation}
where $m_0$ = 0.1\,M$_\odot$, $m_1$ = 0.5\,M$_\odot$, and  $m_2$ = 50\,M$_\odot$. Our choice of $m_2$ is due to our adoption of an ONC-like cluster to simulate; the most 
massive star in the ONC, $\theta^1$\,Ori\,C has a system mass of 45--50\,M$_\odot$ \citep{Kraus07,Kraus09}. We do not include brown dwarfs in the simulations. We do not include primordial binaries, because they do not affect the determination of $\Sigma$.
 
Clusters are unlikely to remain gravitationally bound if they undergo an early phase of gas expulsion\footnote{We note that the recent study by \citet{Kruijssen12} finds that gas expulsion may 
not have a strong effect on the evolution of clusters in hydrodynamical simulations; see also the discussion of recent observational evidence on the relative importance of gas expulsion in \citet{Bastian11}.} \citep{Tutukov78,Lada84,Goodwin06,Bastian08}, and we model this effect by setting half of our simulations to be supervirial, which causes them to 
immediately begin to expand. We also set up clusters in virial equilibrium (static), and subvirial (collapsing).

\subsubsection{Plummer sphere clusters}

The stars have positions and velocities chosen 
according to the prescription from a Plummer sphere in \citet*{Aarseth74}. We set up clusters in two different `states'; one has a half-mass radius, $r_{1/2} = 0.8$\,pc and is in virial equilibrium 
at the start of the simulations (a virial ratio of $Q_{\rm vir} = 0.5$, where $Q_{\rm vir} = T/|\Omega|$ and $T$ and $\Omega$ are the total kinetic and potential energy of the stars, respectively). The second cluster 
has a much smaller half-mass radius ($r_{1/2} = 0.1$\,pc) and is initially supervirial (i.e.\,\,initially expanding), with $Q_{\rm vir} = 1.5$). 

\subsubsection{Fractal clusters}

We set up the fractals with an initial radius of 1\,pc according to the method in \citet{Goodwin04a}. As detailed in Section~\ref{synthetic}, the level of substructure in a fractal is determined by just one number, the fractal dimension, $D$. 
We set up highly substructured fractals, with $D = 1.6$. As with the Plummer sphere clusters, we draw 1000 stars from the IMF in Equation~\ref{imf} and distribute them randomly in the fractal. The velocities of  
the `parents' in the fractal (see Section~\ref{fractal}) are drawn from a Gaussian of mean zero, and the `children' inherit their parent's velocity plus a random component that decreases with each generation of the fractal. 
This means that nearby stars have similar velocities to their neighbours, but distant stars can have very different velocities\footnote{A discussion of the potential effects of non-correlated velocities on the dynamical evolution of the fractal clusters is provided in \citet{Goodwin04a}.}.

Finally, we scale the velocities of the stars in the fractal to the desired virial ratio; in one set of simulations we choose a subvirial (cool) virial ratio, $Q_{\rm vir} = 0.3$, and in the second set of simulations we again adopt a supervirial ratio ($Q_{\rm vir} = 1.5$). \\

We run 10 suites of each simulation, identical apart from the random number seed used to initialise the positions, masses and velocities of the stars. The simulations are evolved for 10\,Myr using the \texttt{kira} integrator in the Starlab package 
\citep[e.g.][]{Zwart99,Zwart01}. We do not include stellar evolution in the simulations. Details of the four simulations are given in Table~\ref{simulations}.

\begin{table}
\caption[bf]{A summary of the different cluster properties adopted for the $N$-body simulations.
The values in the columns are: the number of stars in each cluster ($N_{\rm stars}$), the morphology of the cluster (either a Plummer sphere or fractal), the initial virial ratio of the cluster ($Q_{\rm vir}$), the initial radius of the fractal, ($r_{\rm F}$), or the initial half-mass radius of the Plummer sphere, ($r_{1/2}$).}
\begin{center}
\begin{tabular}{|c|c|c|c|}
\hline 
$N_{\rm stars}$  & Morphology & $Q_{\rm vir}$ &  $r_{\rm F}$ or $r_{1/2}$ \\
\hline
1000 & Plummer sphere & 0.5 & 0.8\,pc \\
1000 & Plummer sphere & 1.5 & 0.1\,pc \\
\hline
1000 & Fractal & 0.3 & 1\,pc \\
1000 & Fractal & 1.5 & 1\,pc \\
\hline
\end{tabular}
\end{center}
\label{simulations}
\end{table}

\section{Results}
\label{results}

In this section we show the $\Sigma$ distributions for various cluster morphologies, and calculate the $\mathcal{Q}$-parameter (in 2D), before showing the effects of dynamical evolution in radially smooth, and substructured clusters on 
both the $\Sigma$ distribution and the $\mathcal{Q}$-parameter.

\subsection{Static clusters}
\label{static_results}

For each of the four morphologies, we perform a  two-sample Kolmogorov-Smirnov (KS) test between the $\Sigma$ distribution for the model cluster and the observed distribution from \citet{Bressert10}. If the test returns a p--value of 
less than 0.01 (arbitrarily chosen to be ``low'') we reject the null hypothesis that there is no difference between the two distributions.  A summary of the KS p--values for all four morphologies is given in Table~\ref{KS_tests}, 
along with the level of structure measured by the $\mathcal{Q}$-parameter.

\begin{figure*}
  \begin{center}
\setlength{\subfigcapskip}{10pt}
\hspace*{-1.5cm}\subfigure[]{\label{Plummer-a}\rotatebox{270}{\includegraphics[scale=0.35]{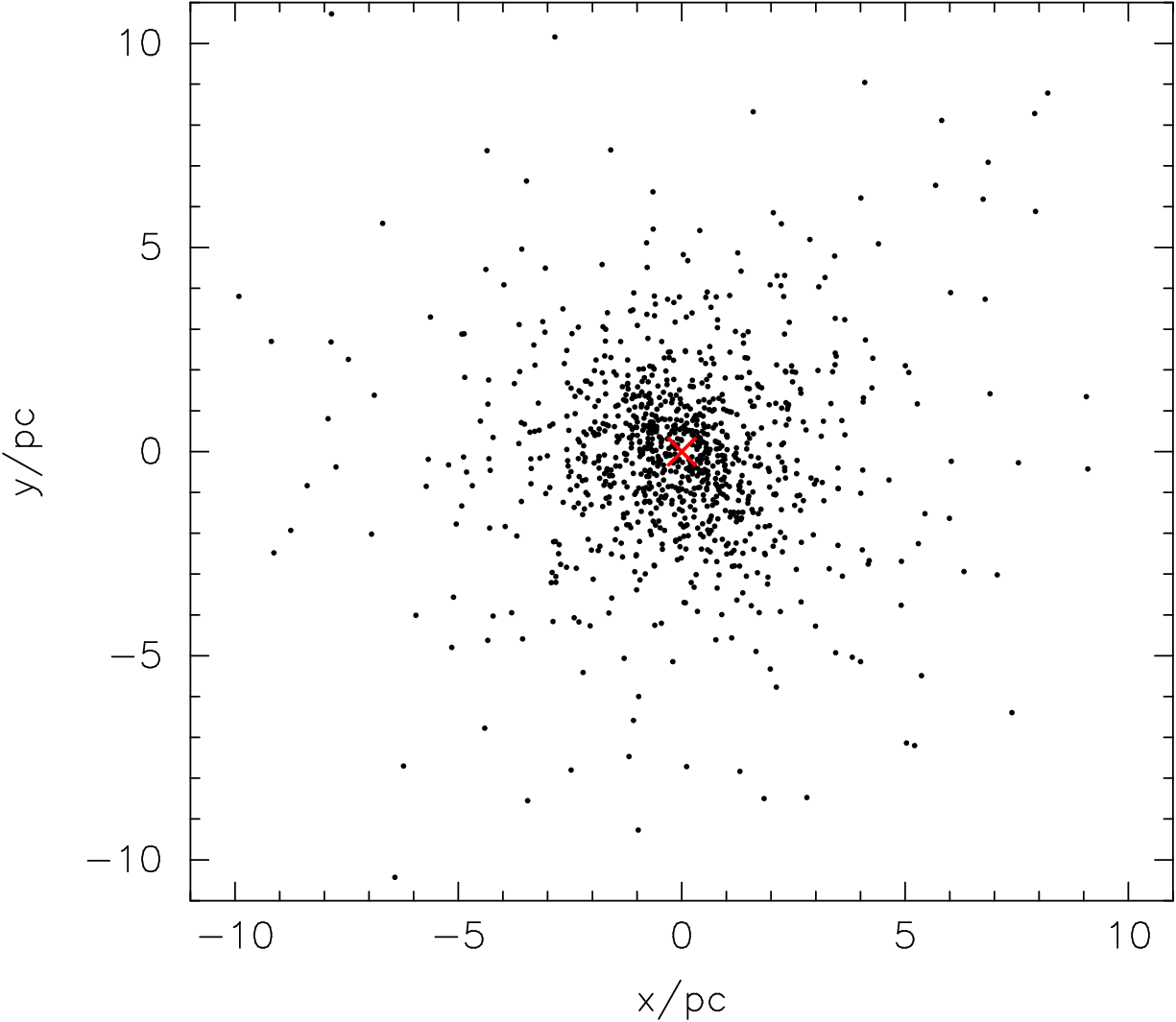}}} 
\hspace*{0.3cm} 
\subfigure[]{\label{Plummer-b}\rotatebox{270}{\includegraphics[scale=0.35]{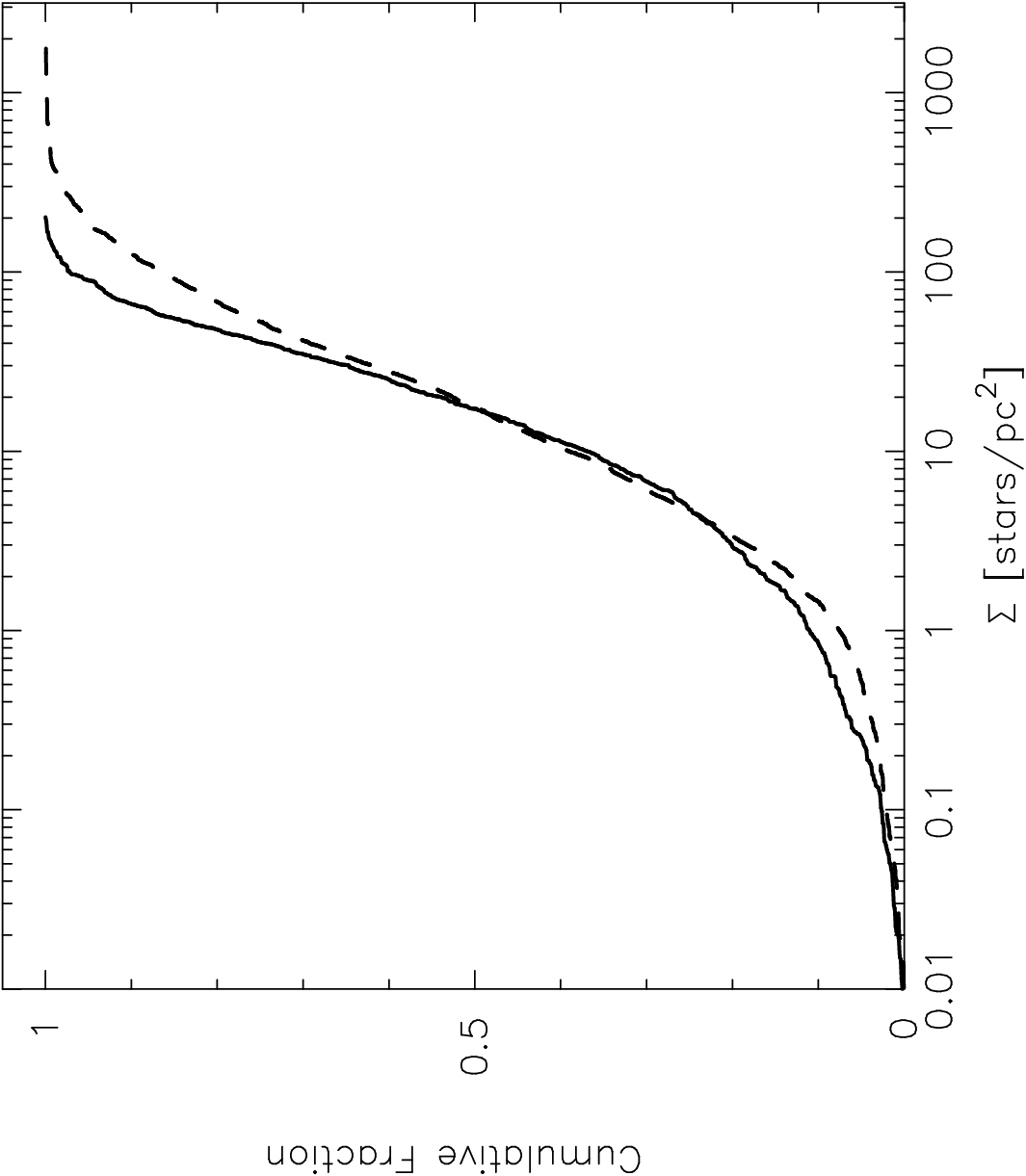}}}
\caption[bf]{A Plummer sphere with half-mass radius $r_{1/2} = 2.0$\,pc, and the corresponding cumulative distribution of stellar surface densities, $\Sigma$. The $\Sigma$ distribution 
is shown by the solid line, and for comparison we show the data from \citet{Bressert10} with the dashed line. The value of $r_{1/2}$ was chosen such that the median $\Sigma$ corresponds to the median value in the observational data ($\tilde{\Sigma}_{\rm YSO} = 22$\,stars\,pc$^{-2}$).}
\label{Plummer}
  \end{center}
\end{figure*}

\begin{figure*}
  \begin{center}
\setlength{\subfigcapskip}{10pt}
\hspace*{-1.5cm}\subfigure[]{\label{King-a}\rotatebox{270}{\includegraphics[scale=0.35]{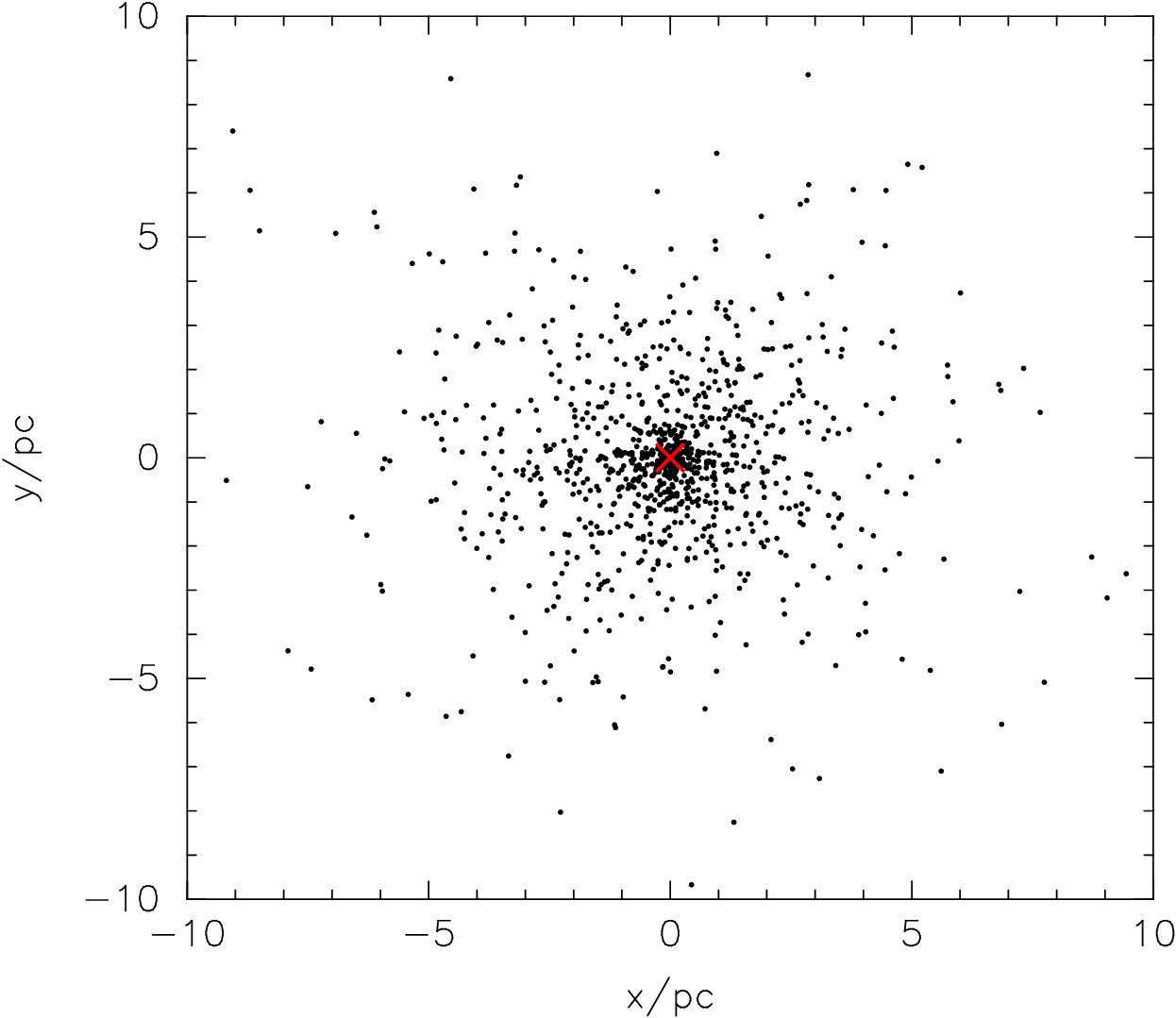}}} 
\hspace*{0.3cm} 
\subfigure[]{\label{King-b}\rotatebox{270}{\includegraphics[scale=0.35]{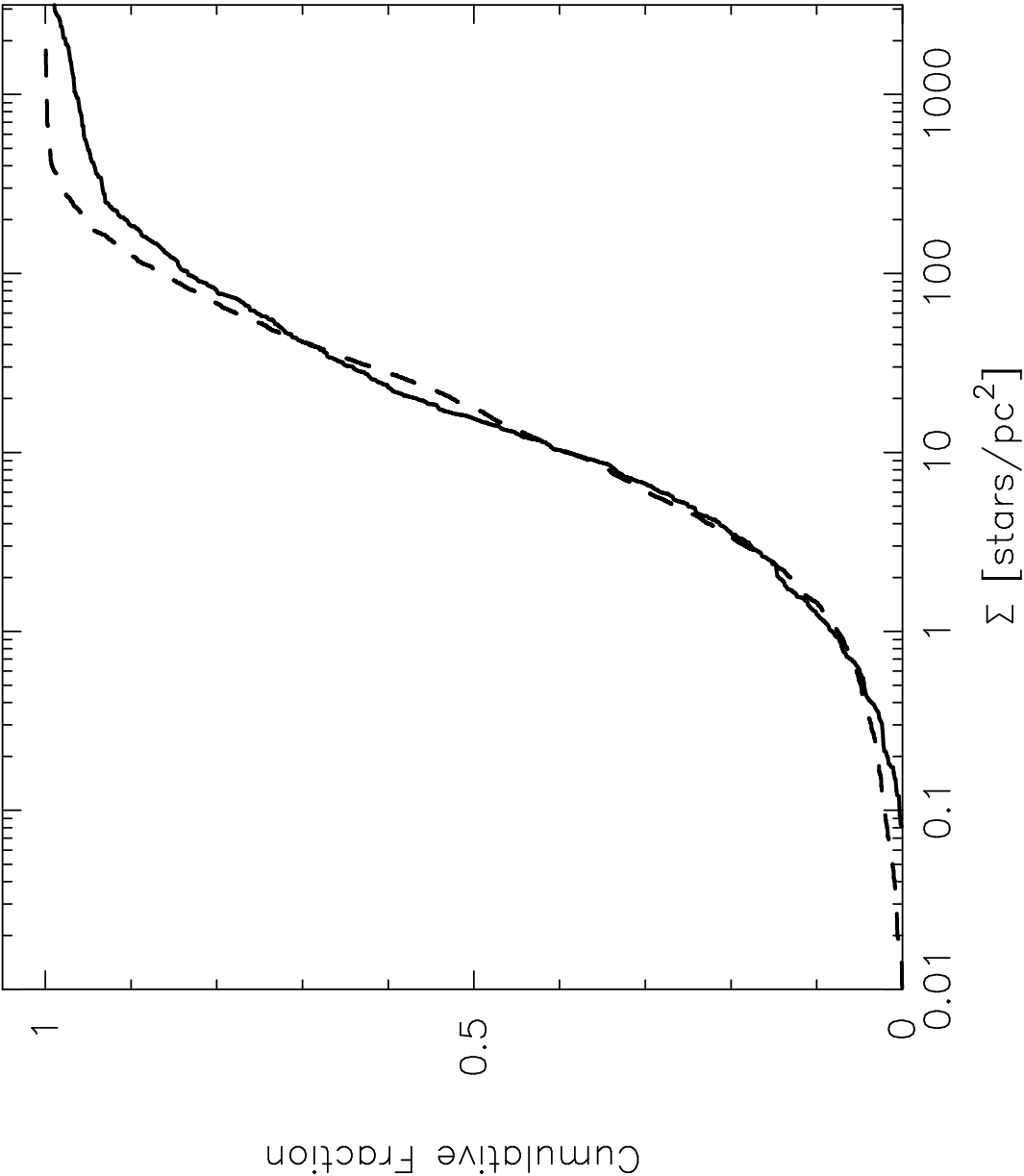}}} 
\caption[bf]{A King profile cluster, with a radius, $r_t$ of 10\,pc and a concentration parameter, $W_0 = 12$,  and the corresponding cumulative distribution of stellar surface densities, $\Sigma$. 
The $\Sigma$ distribution is shown by the solid line, and for comparison we show the data from \citet{Bressert10} with the dashed line. The values of $r_t$ and $W_0$ were chosen such that the median $\Sigma$ corresponds to the median value in the observational data ($\tilde{\Sigma}_{\rm YSO} = 22$\,stars\,pc$^{-2}$). }
\label{King}
  \end{center}
\end{figure*}

\subsubsection{Plummer spheres}

A Plummer sphere with half-mass radius 2.0\,pc is shown in Fig.~\ref{Plummer-a}, with the corresponding 
$\Sigma$ distribution shown in Fig.~\ref{Plummer-b}. At first glance, the Plummer sphere appears to provide a reasonable fit to the observed $\Sigma$ distribution in local star forming regions, and in 
particular it reproduces the tail of the distribution well. 

However, a KS test on the Plummer sphere data and the observational data in \citet{Bressert10} returns a p--value of $< 10^{-8}$  and we reject the null hypothesis that the two datasets are drawn from the same distribution. 

The the level of structure measured by the $\mathcal{Q}$-parameter for this Plummer sphere is $\mathcal{Q} = 1.41$, consistent with a smooth, centrally concentrated profile \citep{Cartwright04}.

\subsubsection{King profiles}

A King profile with a radius of 10\,pc and a concentration parameter $W_0 = 12$ \citep[which corresponds to an extended outer envelope,][]{King66} is shown in Fig.~\ref{King}. This shows a better fit to the observational data than the Plummer sphere, with the only deviation 
occurring at high surface densities.

A KS test returns a p--value of 0.02  and for this morphology we cannot reject the null hypothesis  that the two datasets are drawn from the same distribution.

We find that $\mathcal{Q} = 1.14$, indicating that the King profile is not as centrally concentrated as the Plummer sphere. This is due to our adoption of the King concentration parameter  of $W_0 =12$, which produces an extended outer envelope around the cluster core.

\subsubsection{Fractals}
\label{fractals}

\begin{figure*}
  \begin{center}
\setlength{\subfigcapskip}{10pt}
\hspace*{-1.5cm}\subfigure[]{\label{Fractal1p6-a}\rotatebox{270}{\includegraphics[scale=0.35]{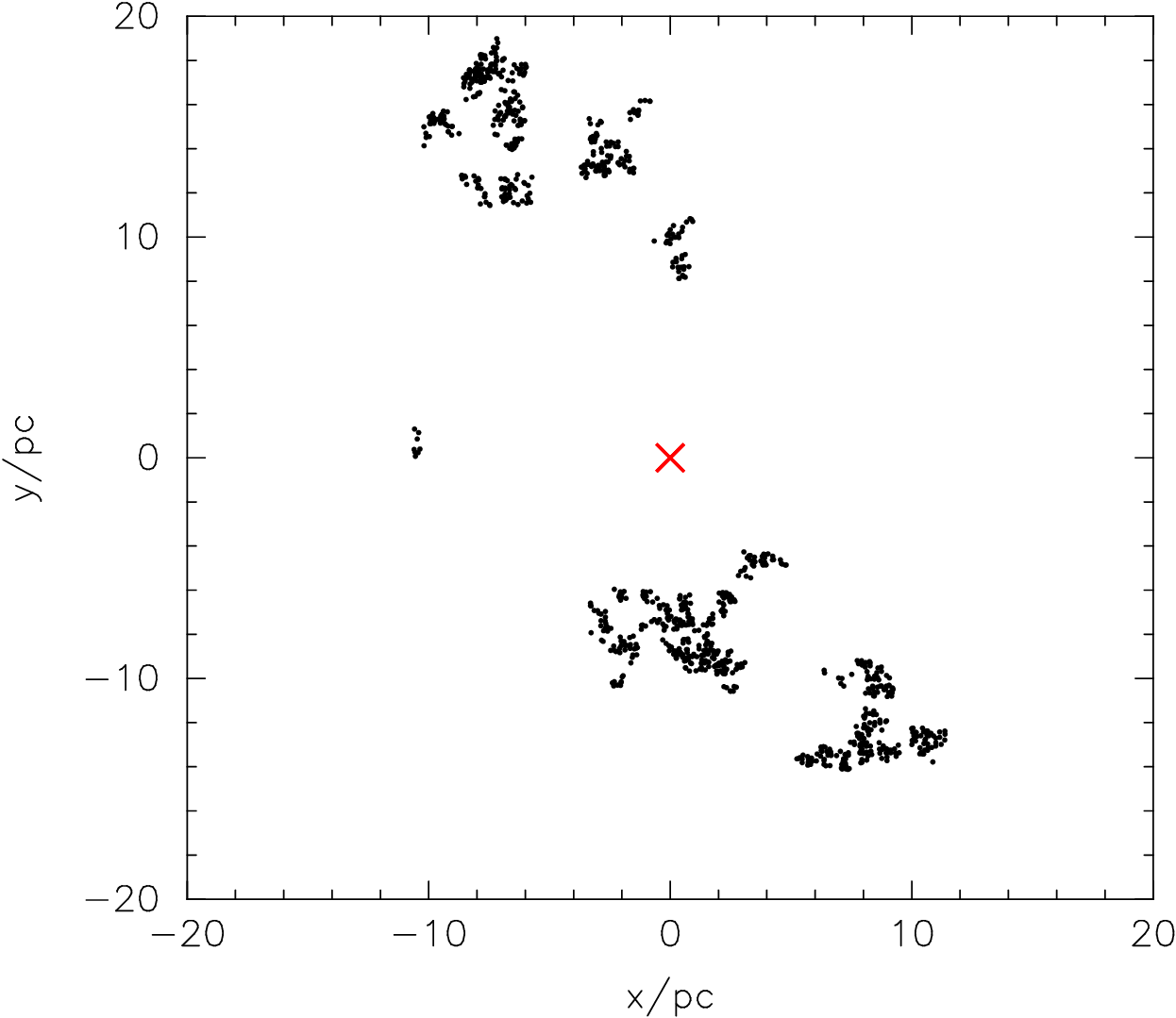}}} 
\hspace*{0.3cm} 
\subfigure[]{\label{Fractal1p6-b}\rotatebox{270}{\includegraphics[scale=0.35]{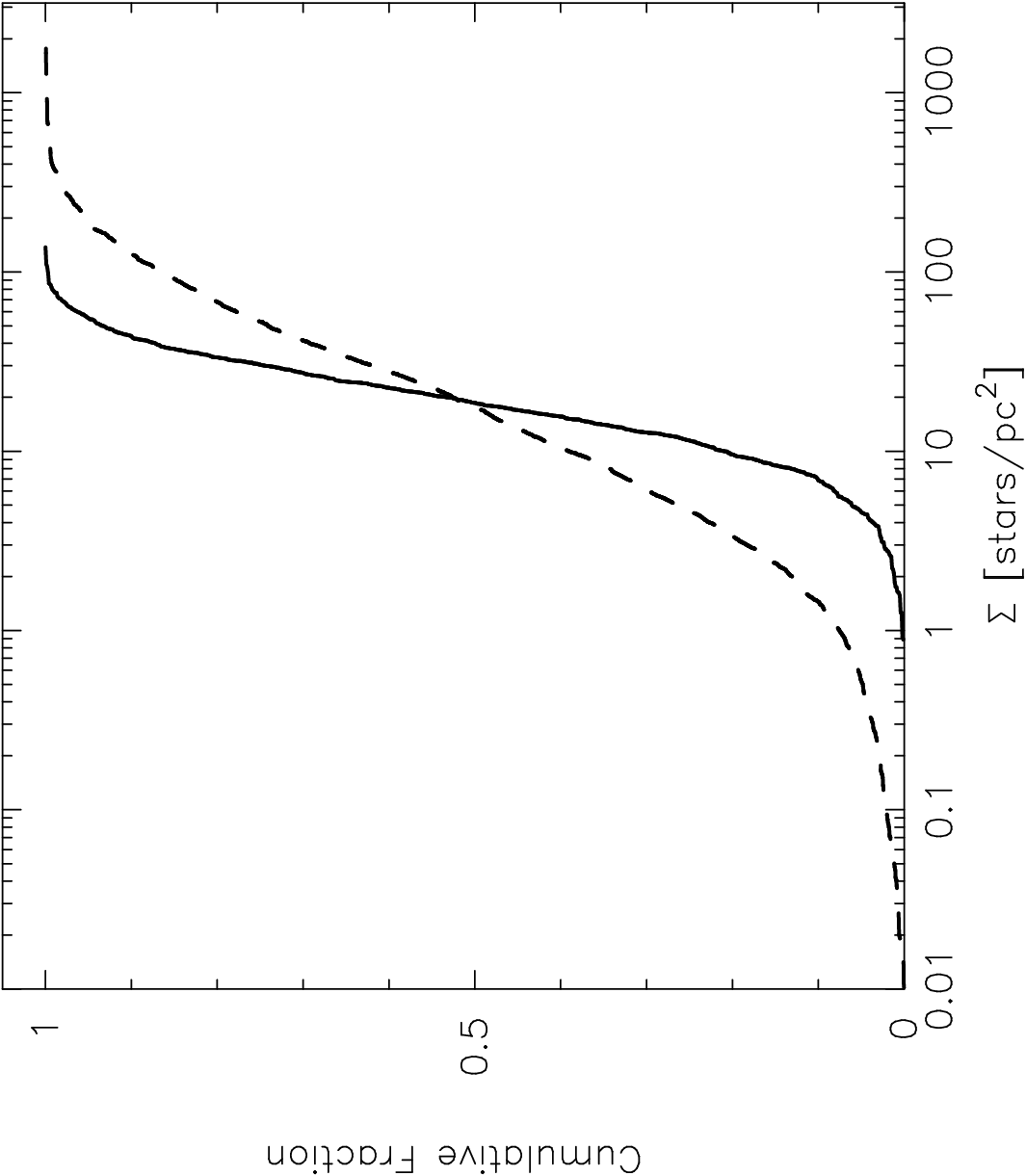}}}
\caption[bf]{A fractal distribution, with fractal dimension $D = 1.6$ and an overall radius of 20\,pc, and the corresponding cumulative distribution of stellar surface densities, $\Sigma$. 
The $\Sigma$ distribution is shown by the solid line, and for comparison we show the data from \citet{Bressert10} with the dashed line. The radius was chosen such that the median $\Sigma$ corresponds to the median value in the observational data ($\tilde{\Sigma}_{\rm YSO} = 22$\,stars\,pc$^{-2}$).}
\label{Fractal1p6}
  \end{center}
\end{figure*}

\begin{figure*}
  \begin{center}
\setlength{\subfigcapskip}{10pt}
\hspace*{-1.5cm}\subfigure[]{\label{Association-a}\rotatebox{270}{\includegraphics[scale=0.35]{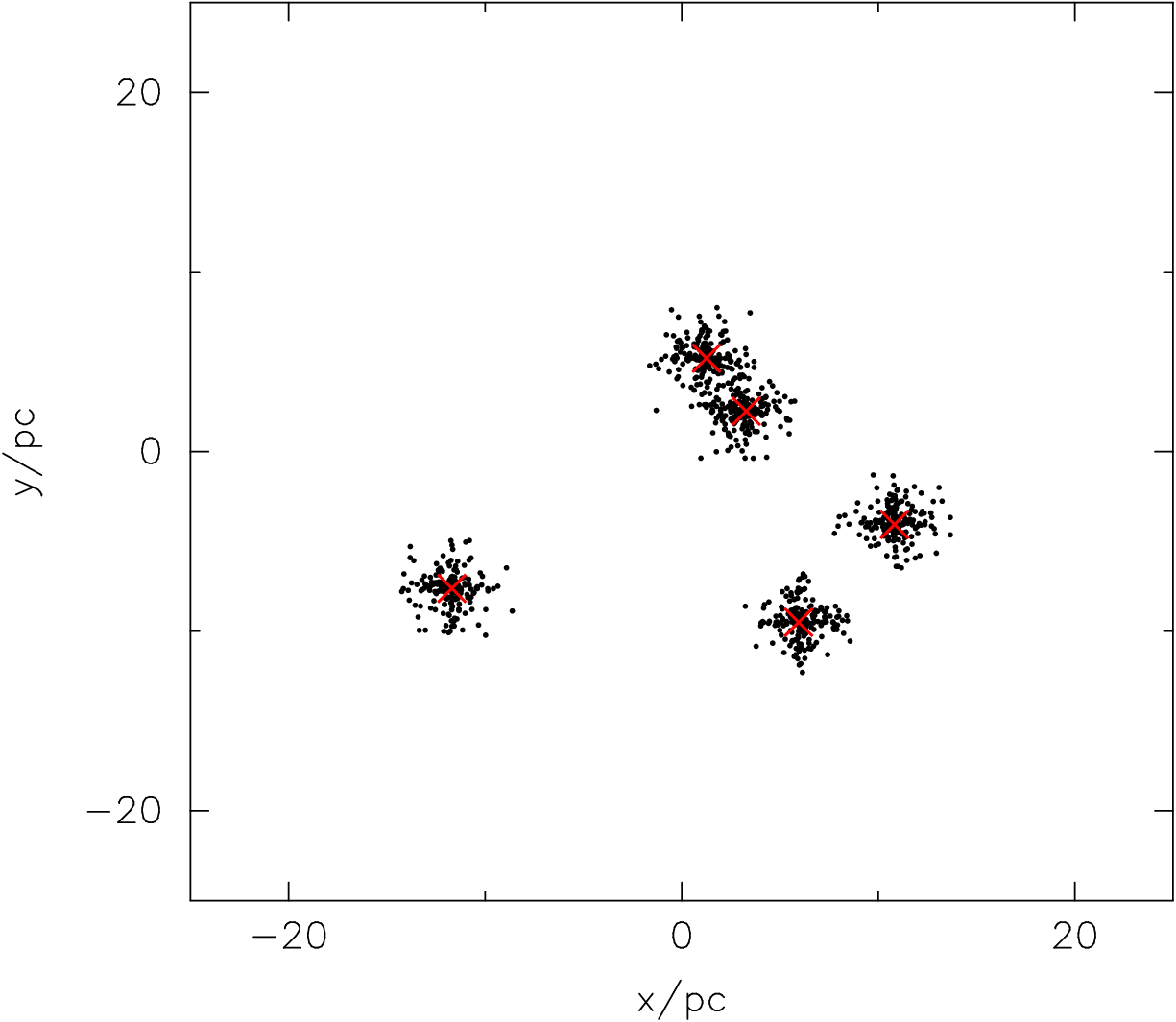}}} 
\hspace*{0.3cm} 
\subfigure[]{\label{Association-b}\rotatebox{270}{\includegraphics[scale=0.35]{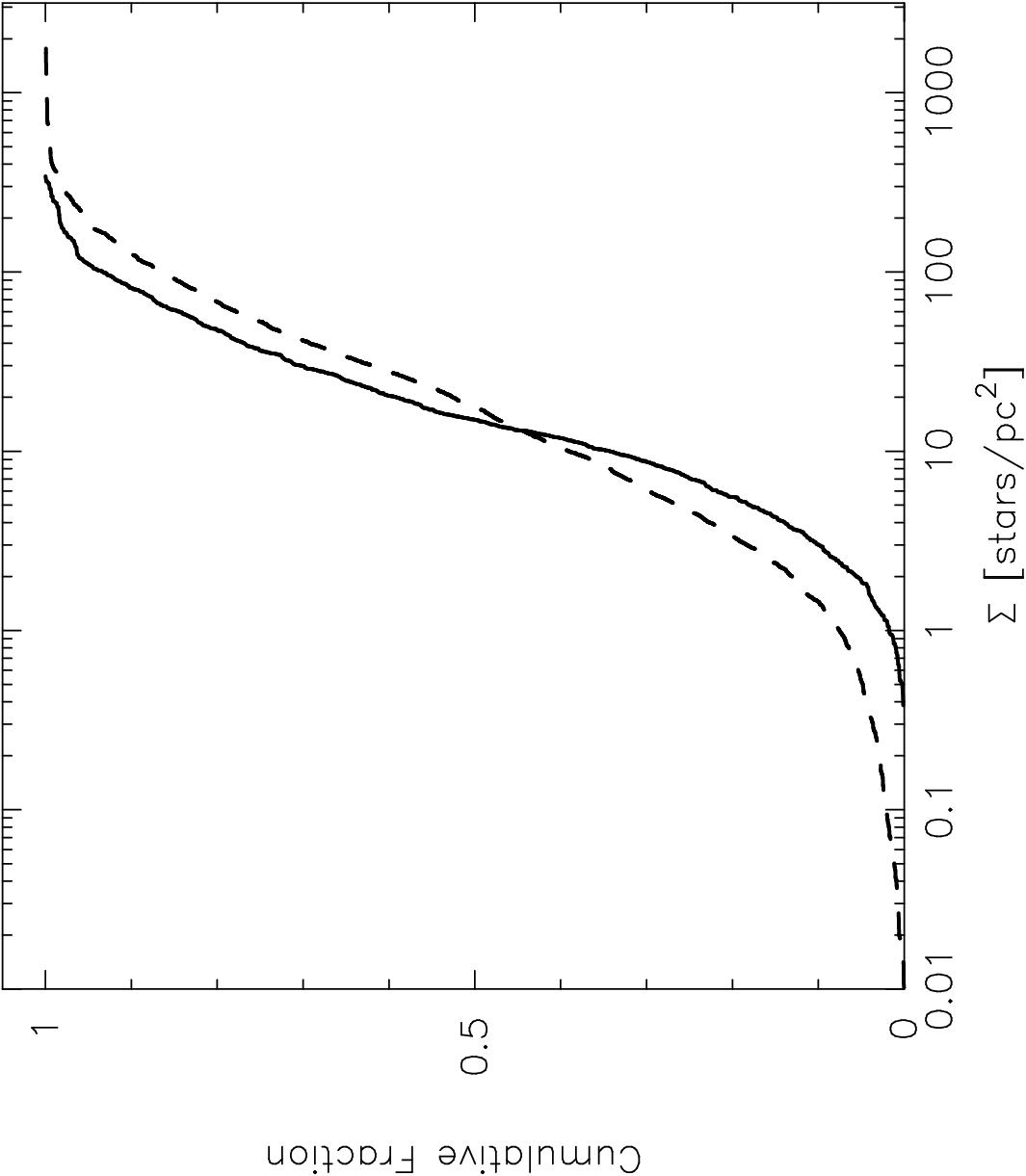}}}
\caption[bf]{An association, comprised of five `nodes' with a cluster radius of 25\,pc, and the corresponding cumulative distribution of stellar surface densities, $\Sigma$. The $\Sigma$ 
distribution is shown by the solid line, and for comparison we show the data from \citet{Bressert10} with the dashed line. The radius was chosen such that the median $\Sigma$ corresponds to the median value in the observational data ($\tilde{\Sigma}_{\rm YSO} = 22$\,stars\,pc$^{-2}$).}
\label{Association}
  \end{center}
\end{figure*}

In Fig.~\ref{Fractal1p6-a} we show a fractal distribution (with fractal dimension $D = 1.6$) and the corresponding $\Sigma$ distribution in Fig.~\ref{Fractal1p6-b}. The fractal has an overall 
scale of 20\,pc. The median value of $\Sigma$ for this fractal agrees with the median value in the \citet{Bressert10} data, but the distribution of $\Sigma$ is much narrower. 

The KS p--value is $< 10^{-35}$  and we reject the null hypothesis that these datasets could be drawn from the same distribution.

The $\mathcal{Q}$-parameter for this fractal is $\mathcal{Q} = 0.20$, indicating a high level of substructure.

\begin{figure*}
  \begin{center}
\setlength{\subfigcapskip}{10pt}
\hspace*{-1.5cm}\subfigure[]{\label{bg-a}\rotatebox{270}{\includegraphics[scale=0.25]{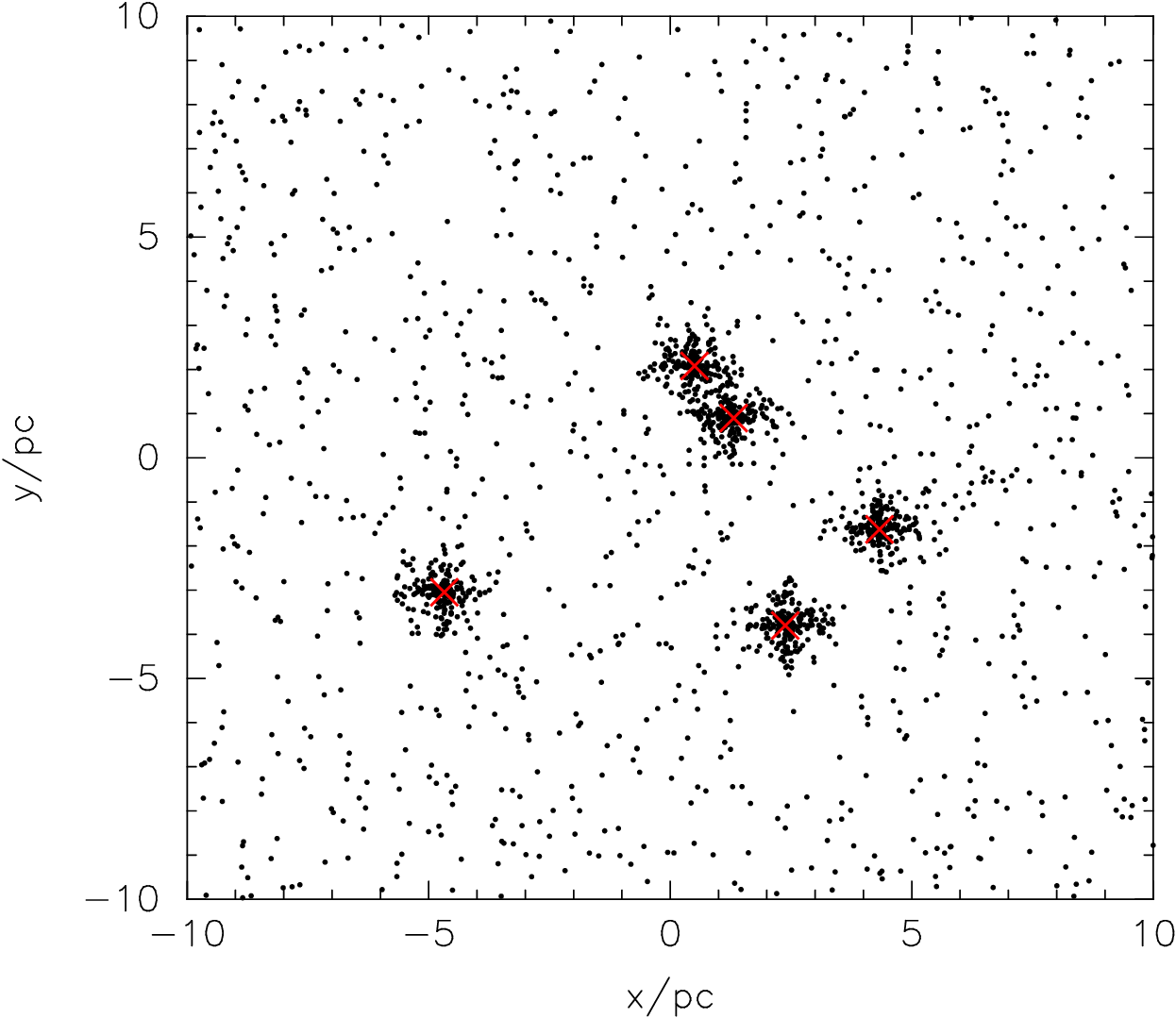}}} 
\hspace*{0.3cm} 
\subfigure[]{\label{bg-b}\rotatebox{270}{\includegraphics[scale=0.25]{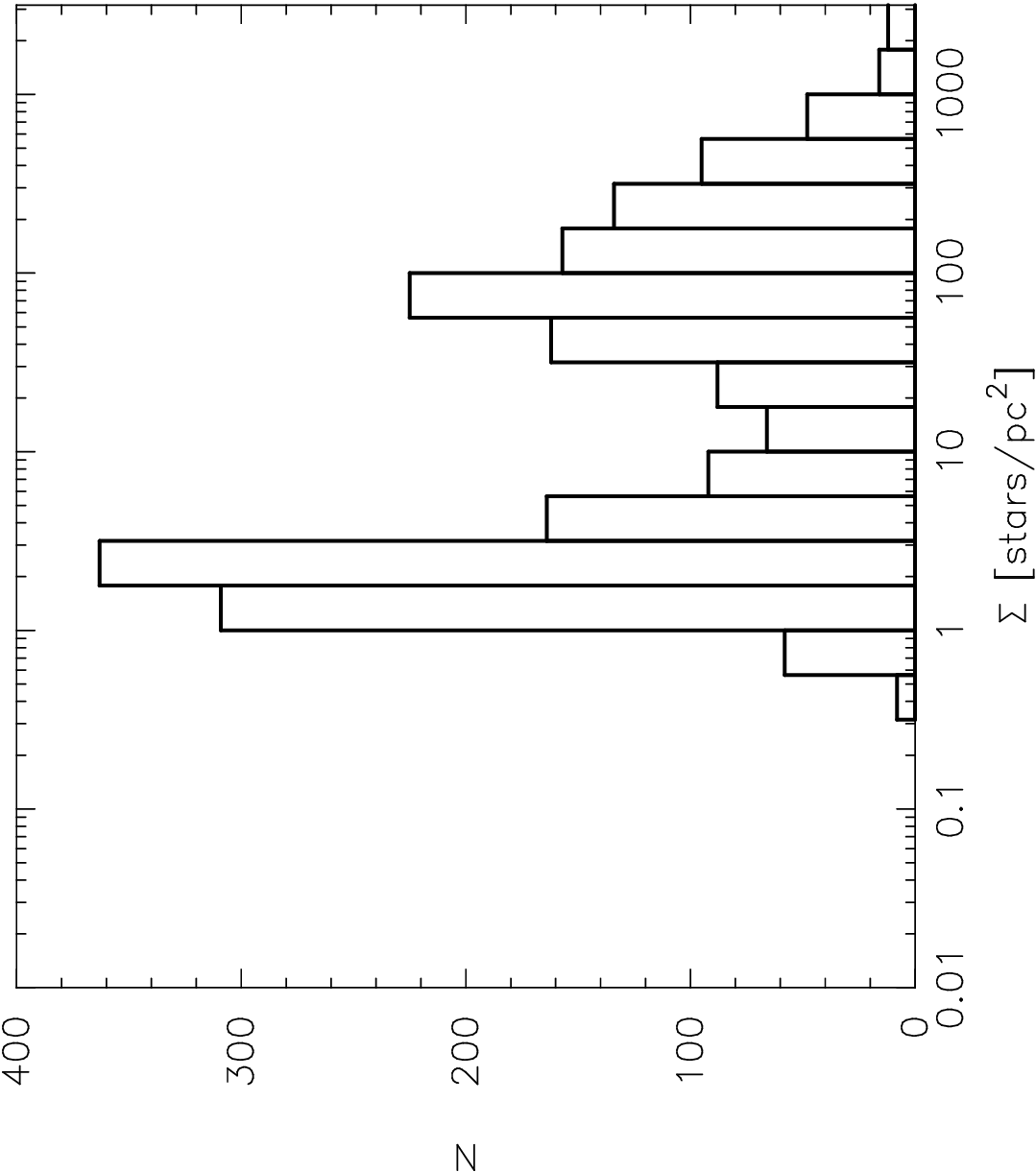}}} 
\hspace*{0.3cm} 
\subfigure[]{\label{bg-c}\rotatebox{270}{\includegraphics[scale=0.25]{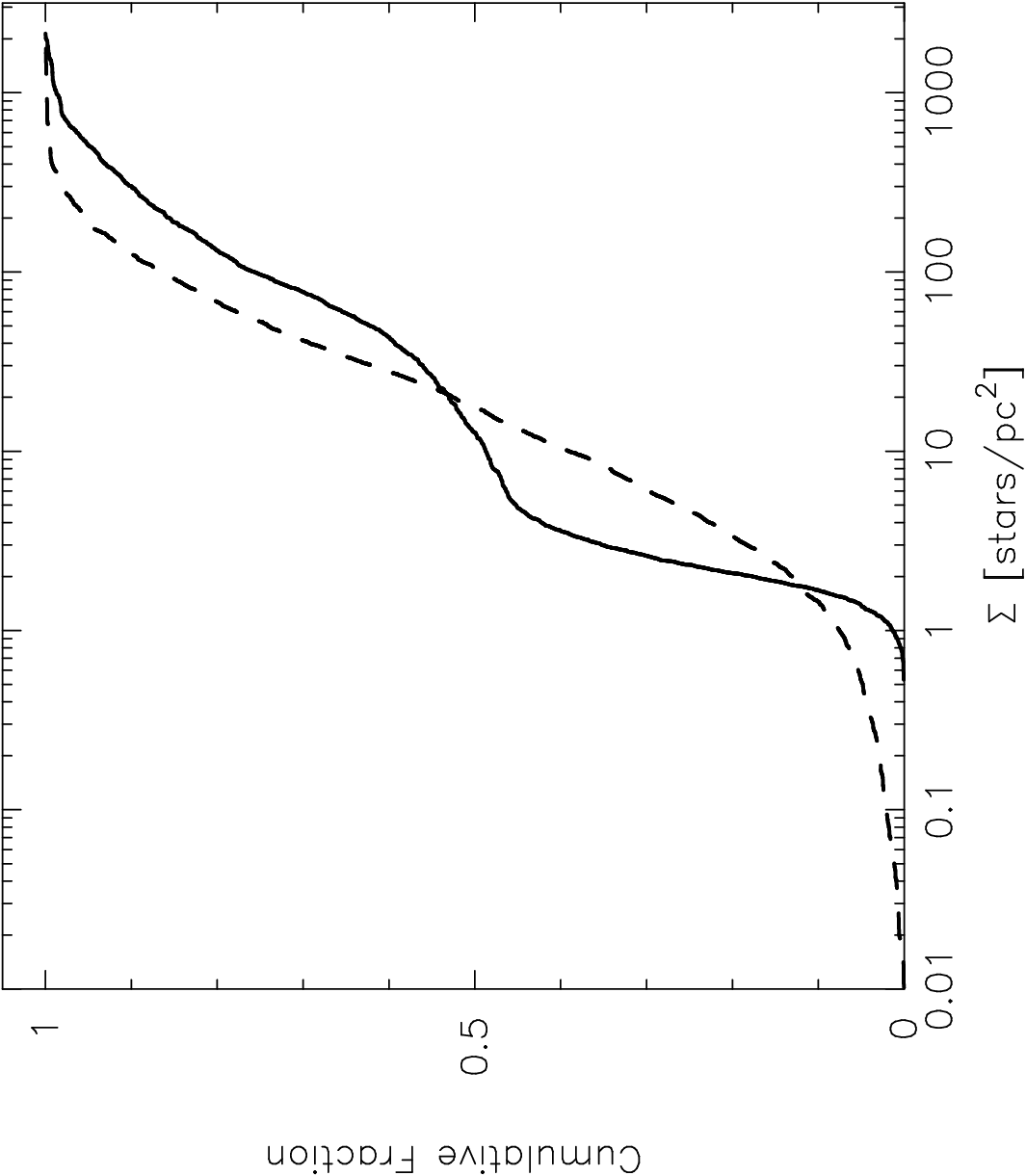}}}
\hspace*{-1.5cm}\subfigure[]{\label{bg-d}\rotatebox{270}{\includegraphics[scale=0.25]{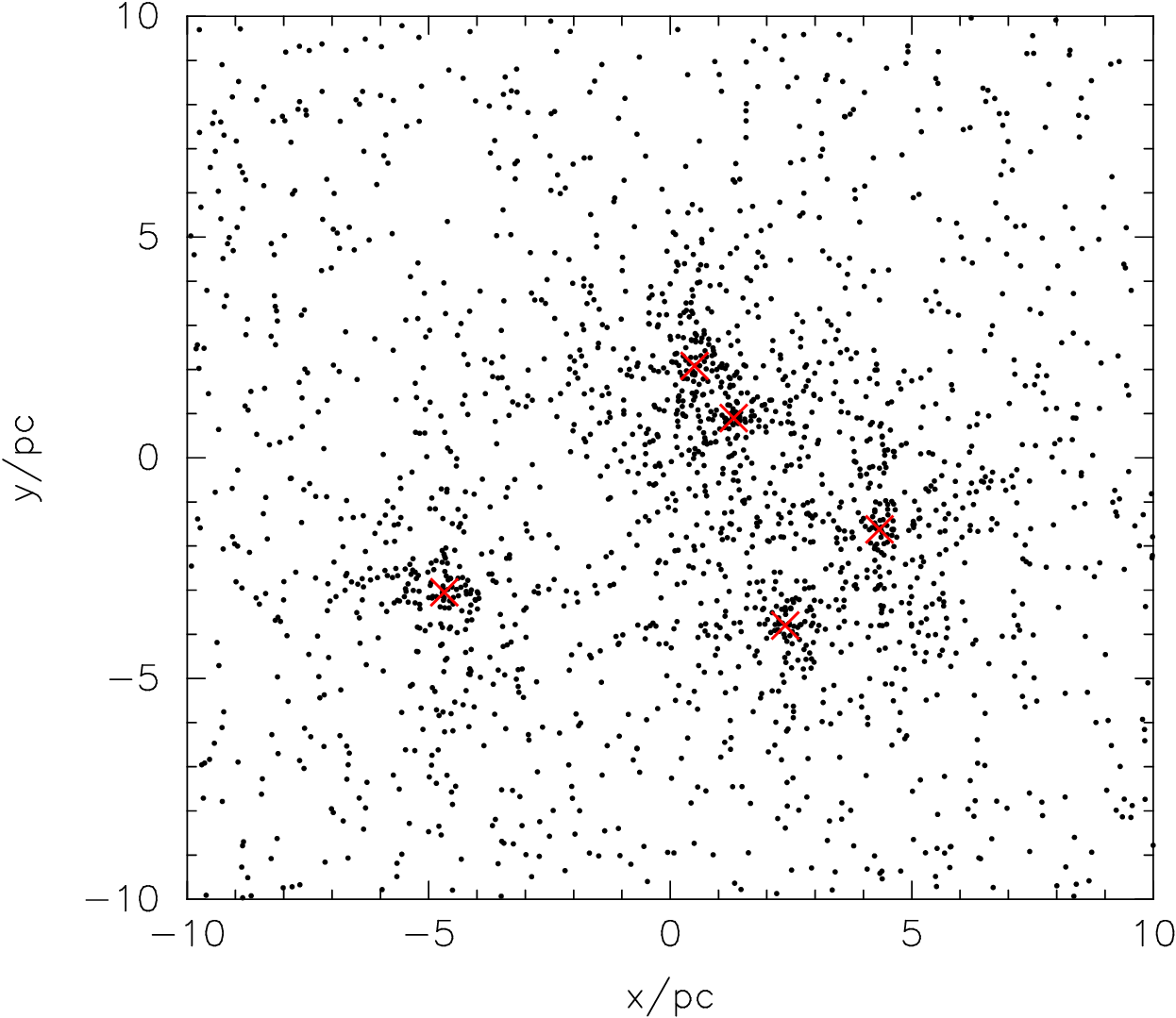}}} 
\hspace*{0.3cm} 
\subfigure[]{\label{bg-e}\rotatebox{270}{\includegraphics[scale=0.25]{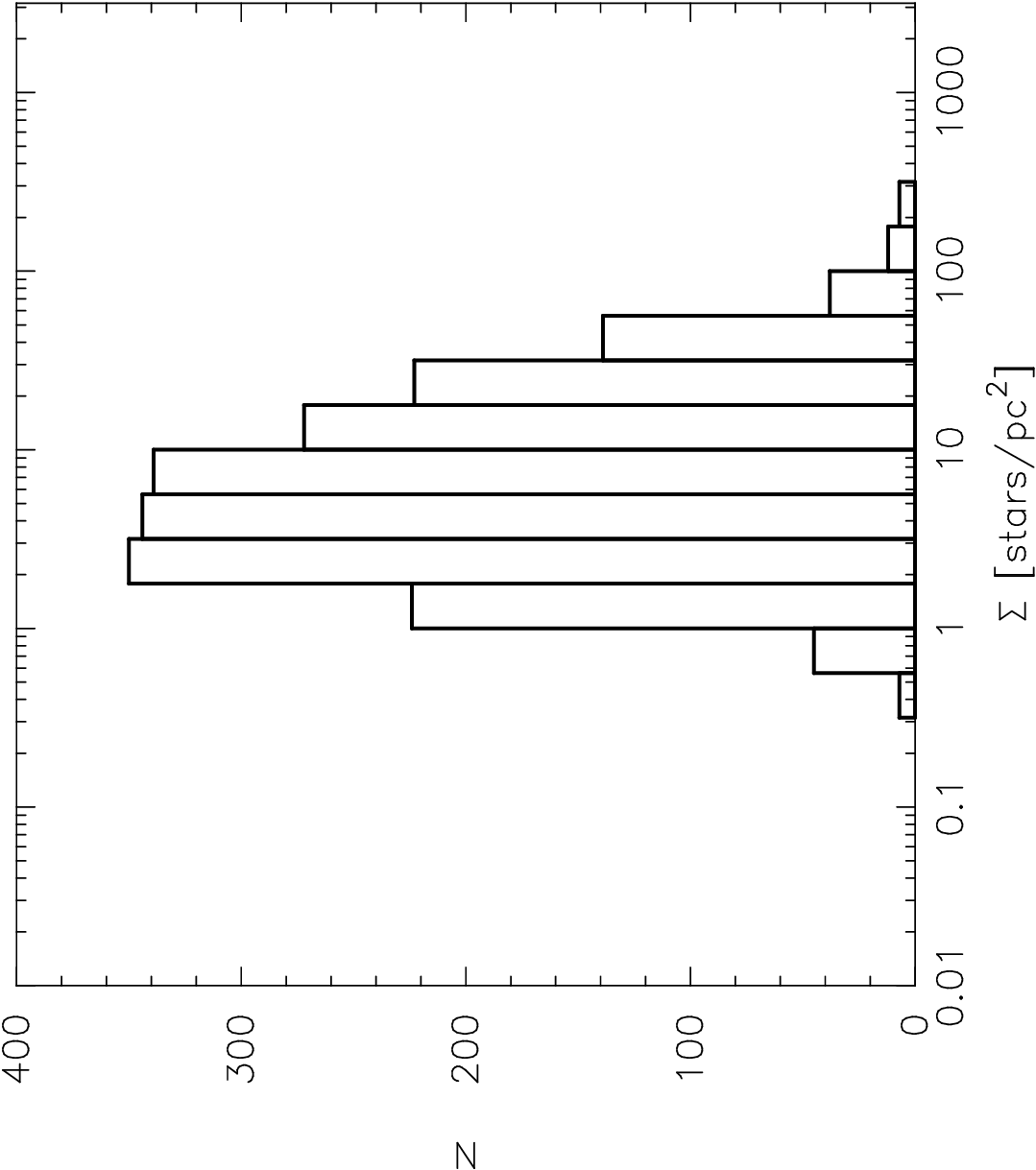}}} 
\hspace*{0.3cm} 
\subfigure[]{\label{bg-f}\rotatebox{270}{\includegraphics[scale=0.25]{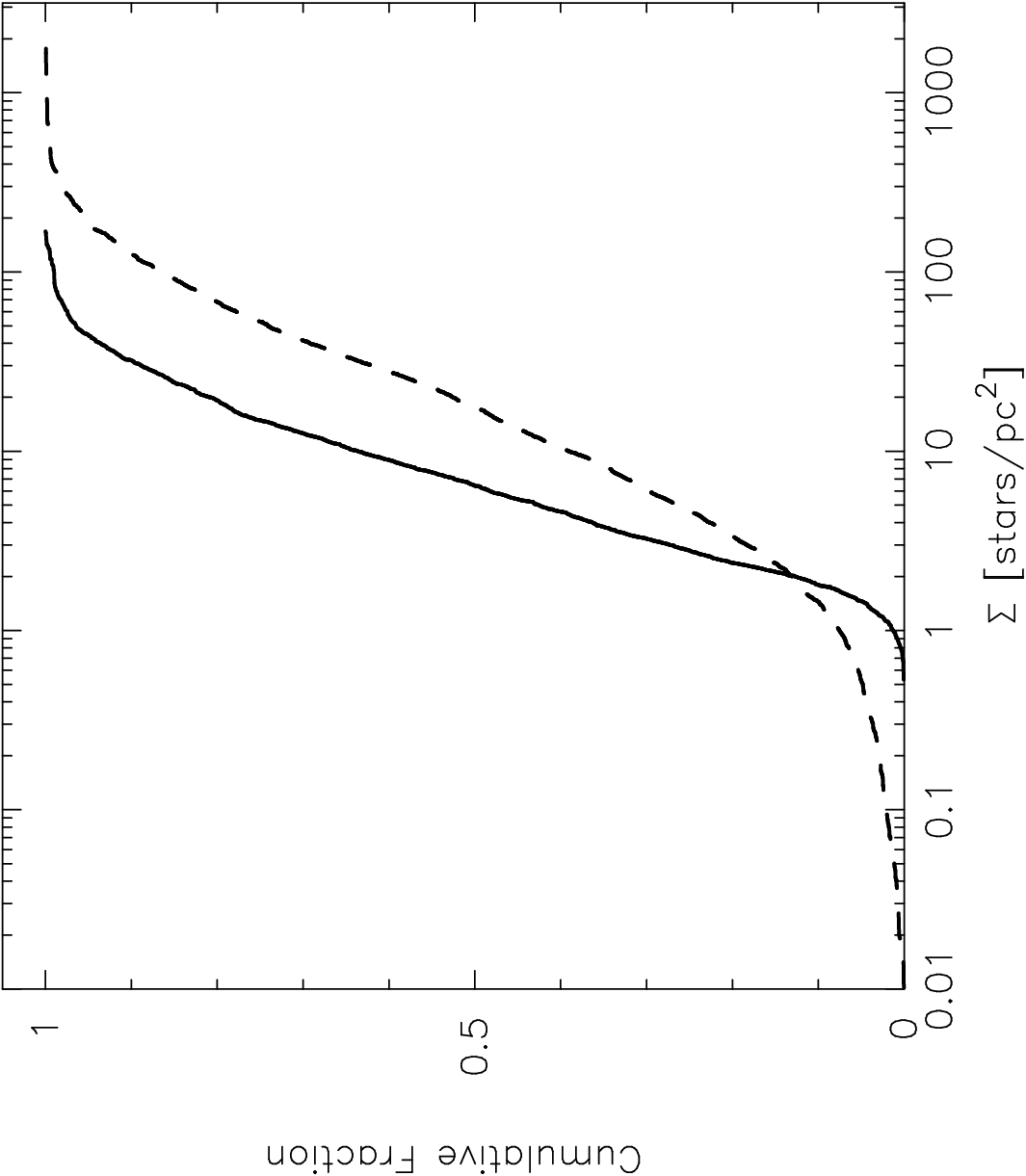}}}
\caption[bf]{The effect of background stars on the distribution of $\Sigma$. In panel (a) we show a cluster with size 10\,pc and a nodal `compactness' $C = 7.5$, against a uniform 
background distribution of 1000 stars. The histogram of $\Sigma$ is shown in panel (b) and the corresponding cumulative distribution of $\Sigma$ is shown in panel (c). The $\Sigma$ distribution is shown by the solid line, and for comparison we 
show the data from \citet{Bressert10} with the dashed line. In panel (d) we reduce the compactness of the association nodes, so that $C = 2.0$, and we retain the spatial density of the background stars. The histogram of $\Sigma$ and the corresponding cumulative distribution are shown in panels (e) and (f), respectively.}
\label{bg}
  \end{center}
\end{figure*}

\subsubsection{Associations} 

In Fig.~\ref{Association} we show an association with radius $R = 25$\,pc, with five nodes placed randomly in three dimensions. The compactness of each node, $C$, is 7.5. As with the 
fractal, the median value of $\Sigma$ agrees well with the mean value from the observational sample; however, and like the fractal, the distribution is not wide enough.  

The KS p--value is $< 10^{-5}$  and we reject the null hypothesis that these datasets could be drawn from the same distribution.

The calculation of the $\mathcal{Q}$-parameter for this association gives $\mathcal{Q} = 0.32$, indicating it is substructured, but not to the same extent as the fractal.

\begin{table}
\caption[bf]{A summary of the four different morphologies of our static clusters. We list the 2D $\mathcal{Q}$-parameter for each morphology, and the KS test p-value between the $\Sigma$ distribution of each cluster and the observational data from \citet{Bressert10}.}
\begin{center}
\begin{tabular}{|c|c|c|}
\hline 
Morphology & KS p--value & $\mathcal{Q}$-parameter \\
\hline
Plummer sphere & $3 \times 10^{-9}$ & 1.41\\
King Profile & $2 \times 10^{-2}$ & 1.14 \\
Fractal & $8 \times 10^{-36}$ & 0.20 \\
Association & $4 \times 10^{-6}$ & 0.32\\
\hline
\end{tabular}
\end{center}
\label{KS_tests}
\end{table}

\subsubsection{Possible biases}

In principle there are several biases that could affect the determination of $\Sigma$ in star forming regions. Firstly, the $\Sigma$ distribution could be affected by the erroneous inclusion of 
foreground or background stars. Depending on the density of the star forming region compared to the background, the inclusion of field stars shows an obvious signature. In Fig.~\ref{bg} 
we show the same association set-up in Fig.~\ref{Association}, with the same compactness, $C = 7.5$ but with a smaller cluster size (10\,pc). We then randomly place a further 1000 stars within the field of view and 
determine the distribution of $\Sigma$ including all stars. The histogram of $\Sigma$ clearly shows bimodality in the distribution (Fig.~\ref{bg-b}) and in Fig.~\ref{bg-c} we show the cumulative distribution, which displays a prominent kink in the transition from the cluster nodes to 
the background. 

We now change the $C$ parameter of the nodes in the association to $C = 2.0$. This has the effect of dispersing the stars in the nodes so that they are less distinct against the 
background (see Fig.~\ref{bg-d}). The bimodality is no longer apparent in the histogram (Fig.~\ref{bg-e}) (although it is still distinguishable by eye) and the cumulative distribution of $\Sigma$ reverts to a smooth line (Fig.~\ref{bg-f}).

\begin{figure*}
  \begin{center}
\setlength{\subfigcapskip}{10pt}
\hspace*{-1.5cm}\subfigure[]{\label{extinct-a}\rotatebox{270}{\includegraphics[scale=0.35]{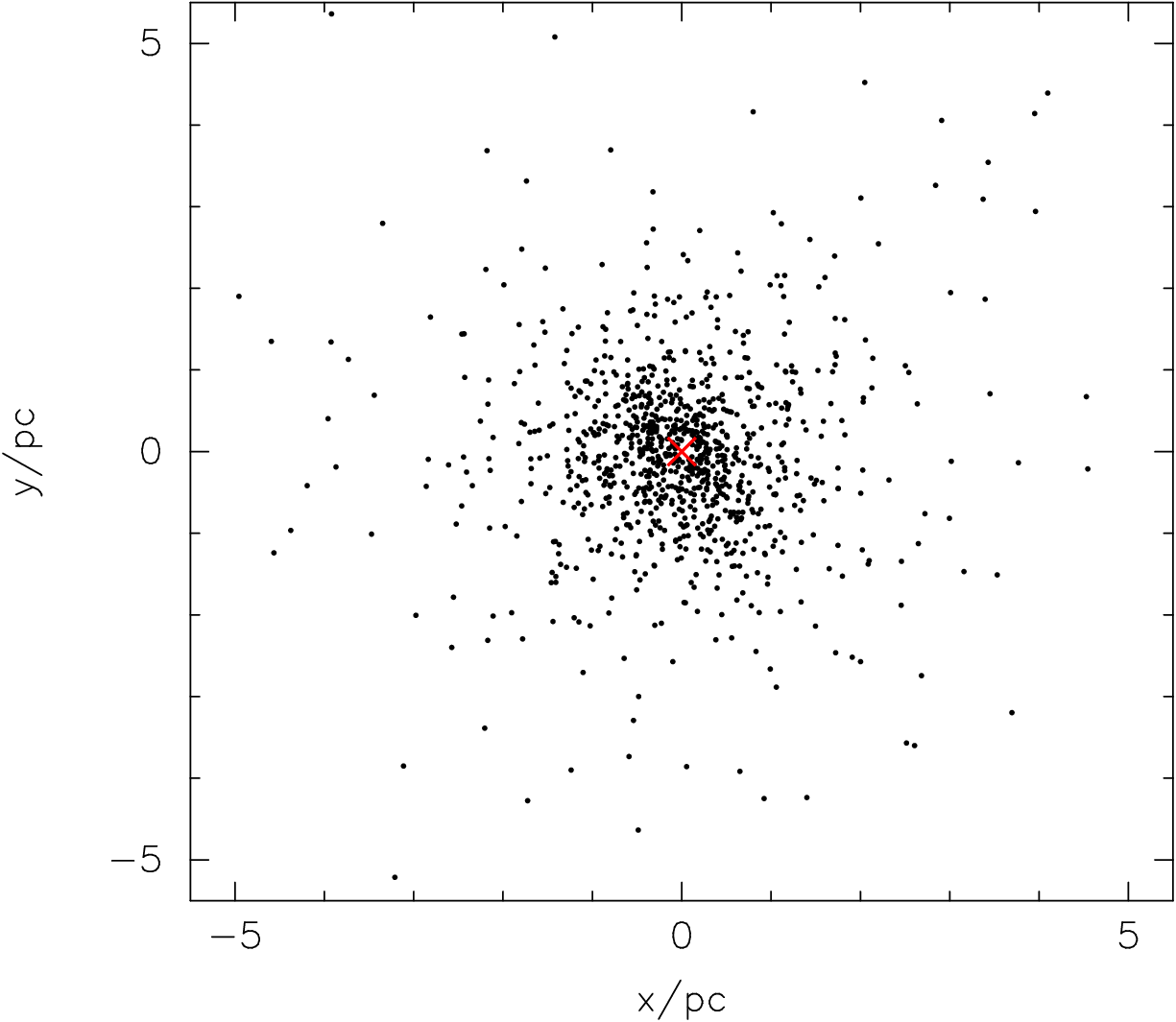}}} 
\hspace*{0.3cm} 
\subfigure[]{\label{extinct-b}\rotatebox{270}{\includegraphics[scale=0.35]{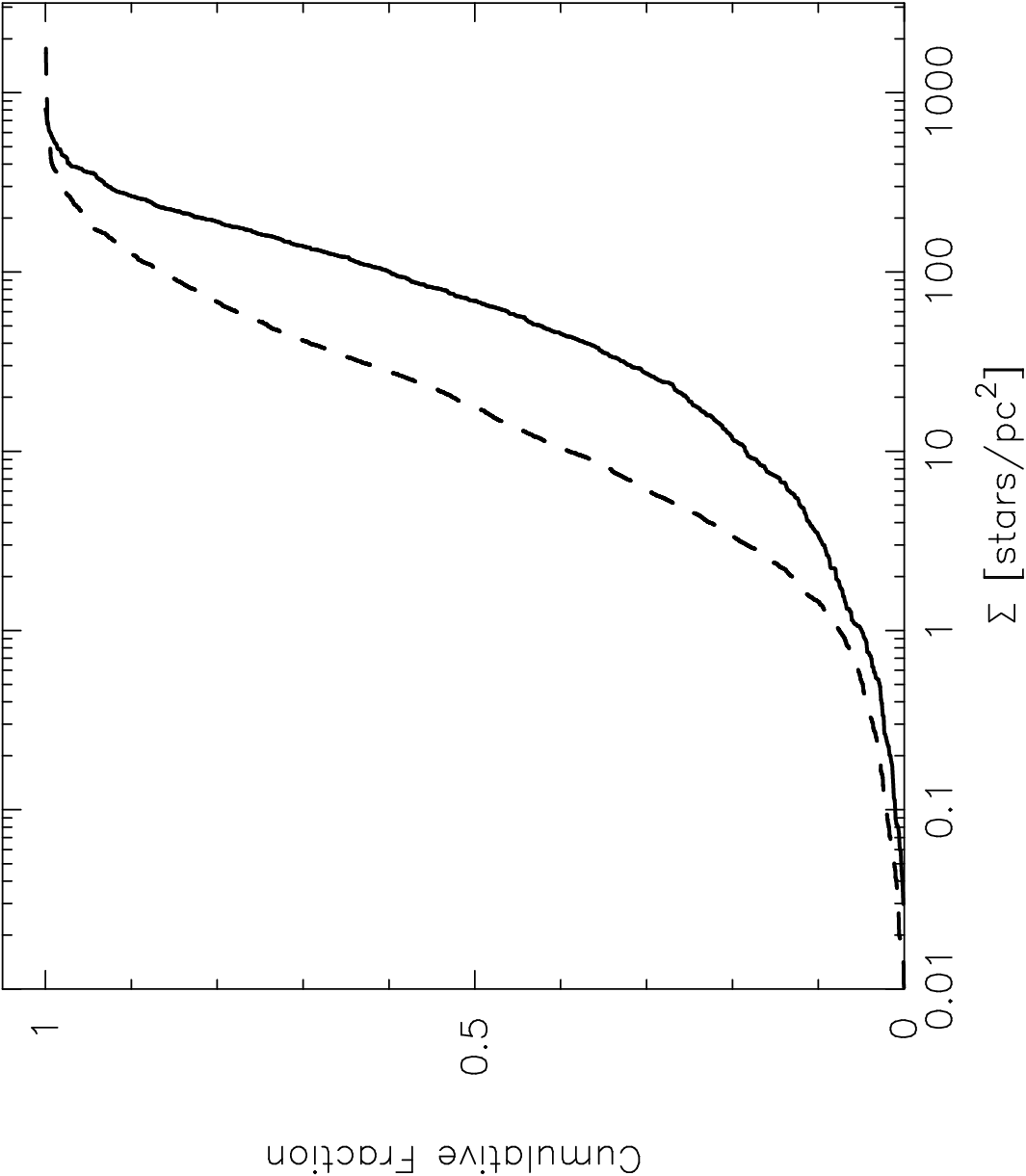}}}
\hspace*{-1.5cm}\subfigure[]{\label{extinct-c}\rotatebox{270}{\includegraphics[scale=0.35]{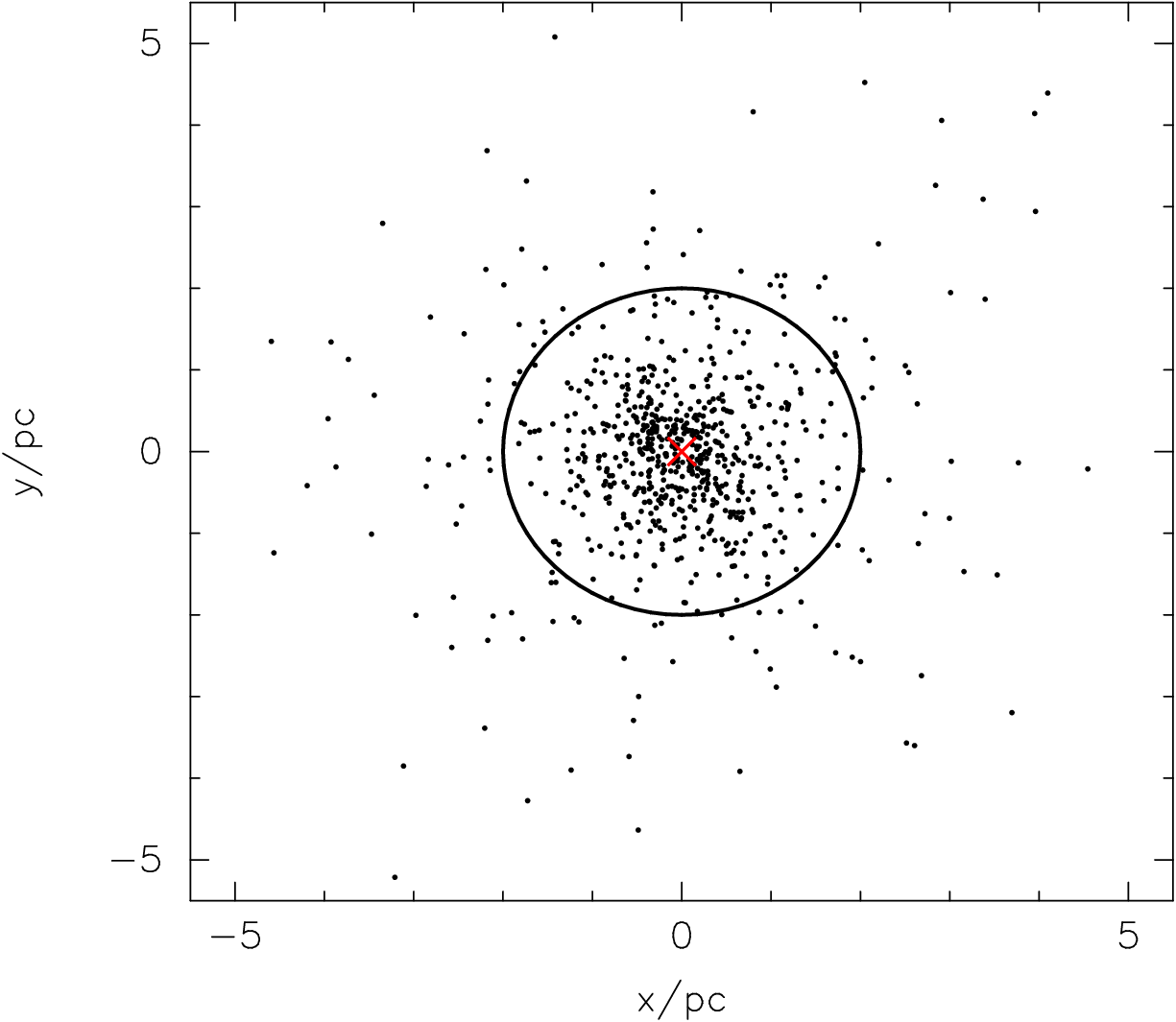}}} 
\hspace*{0.3cm} 
\subfigure[]{\label{extinct-d}\rotatebox{270}{\includegraphics[scale=0.35]{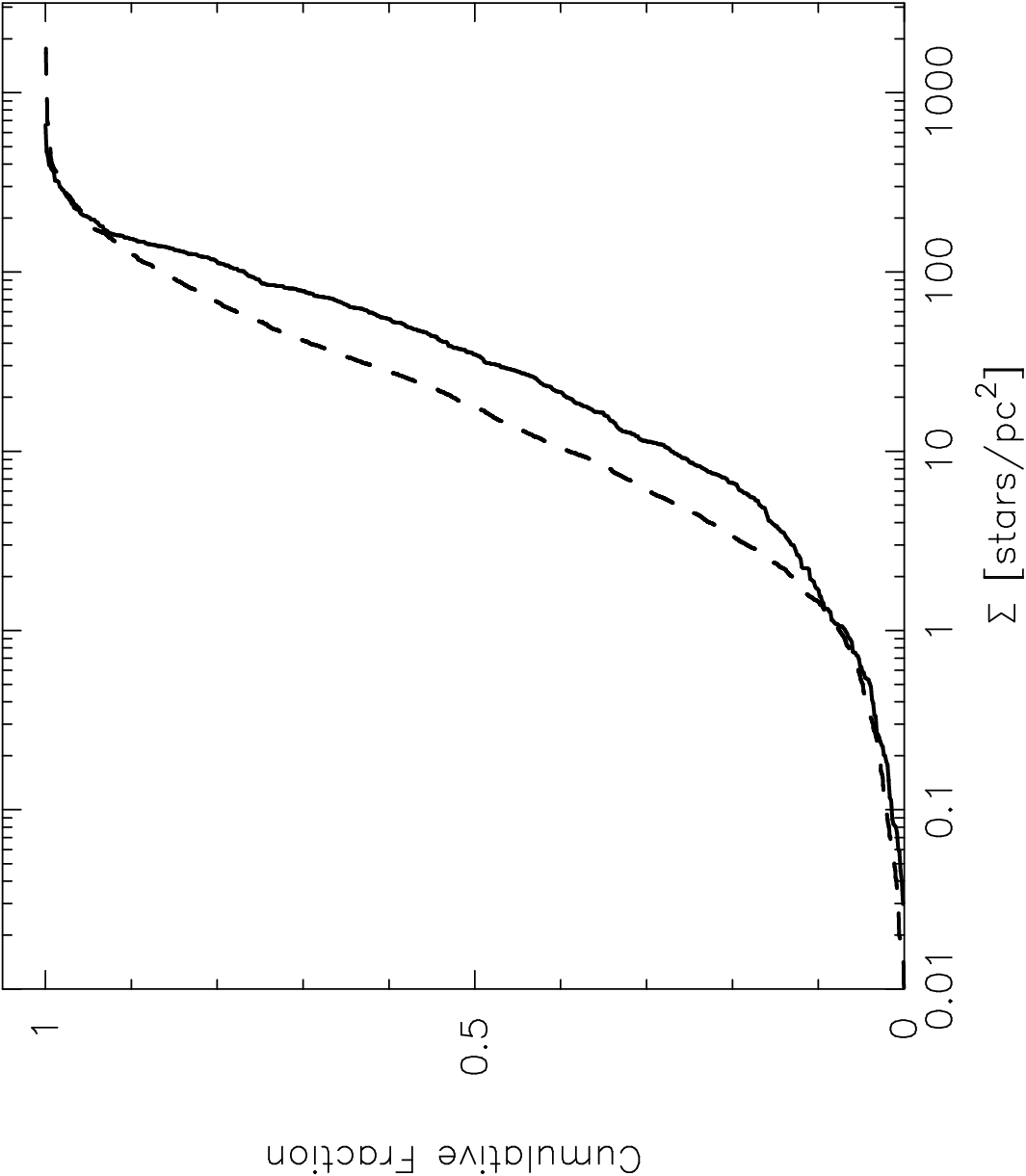}}}
\caption[bf]{Plummer spheres with half-mass radii $r_{1/2} = 1.0$\,pc, and the respective $\Sigma$ cumulative distributions before removal of extincted stars [(a) and (b)], and after  [(c) and (d)]. 
The extent of the dust cloud ($2.0\,r_{1/2}$) is shown by the circle in panel (c). The $\Sigma$ distribution is shown by the solid line, and for comparison we show the data from \citet{Bressert10} 
with the dashed line.}
\label{extinct}
  \end{center}
\end{figure*}

Secondly, the determination of $\Sigma$ could be affected by ``missing'' stars, hidden by dust extinction. We examine the effects of extinction on a flux limited sample for a Plummer sphere cluster by applying a simple extinction power law to the cluster. We reduce the level of extinction from $A_v = 100$ 
at the cluster centre  to $A_v = 0$  at a distance of $2.0\,r_{1/2}$, where $r_{1/2}$ is the half-mass radius, using the following formula:
\begin{equation}
A_v(r) = 100\left[1 - \left(\frac{|r|}{1.5\,r_{1/2}}\right)^5\right],
\end{equation}
where $r$ is the position of the star with respect to the cluster centre and $|r|$ is the modulus of its vector. To account for projection effects along the line of sight, we double $A_v(r)$ if 
the $z$-component of the vector $r$ is negative. This encompasses the densest region of the Plummer sphere (see Fig.~\ref{extinct-c}). We have varied the power-law exponent between 3 and 7 and find 
very little difference to the results.

If a star has a mass $m < 0.2$\,M$_\odot$ and has an extinction 
$A_v(r) > 25$, then it is `hidden', and not counted in the determination of $\Sigma$. We find that 25 per cent of stars are hidden if we apply the above constraints. 

The results of extinction on the distribution of $\Sigma$ are shown in Fig.~\ref{extinct}. In panel (a) we show the original Plummer sphere (with half-mass radius $r_{1/2} = 1.0$\,pc), before the 
extincted stars are removed. In panel (b) we  show the distribution of $\Sigma$ by the solid line, with the data from \citet{Bressert10} shown for comparison. In panel (c) we show the cluster 
with the extincted stars removed (and the extent of the dust cloud shown by the circle) and the corresponding $\Sigma$ distribution in panel (d). The effect of extinction is to move the 
distribution to the left, indicating that a star forming region could appear less dense due to `hidden stars'. However, we note that the level of extinction required to produce this shift is 
drastic, and in practice is unlikely to bias observational studies. 

Similarly, the determination of the $\mathcal{Q}$-parameter is not strongly affected by the removal of extincted stars. For a Plummer sphere we found $\mathcal{Q} = 1.41$, which is reduced to $\mathcal{Q} = 1.36$ 
after the removal of 25\,per cent of the stars in Fig.~\ref{extinct-a}. This is consistent with the findings of \citet{Bastian09}, who find that the $\mathcal{Q}$-parameter decreases by $0.07 \pm 0.03$ from the true  value when stars in a sample 
are hidden by extinction. 

\subsection{Dynamical evolution}

\begin{figure*}
  \begin{center}
\setlength{\subfigcapskip}{10pt}
\hspace*{-1.5cm}\subfigure[$Q_{\rm vir} = 0.5$ (virial equilibrium) ]{\label{plummer_evolve-a}\rotatebox{270}{\includegraphics[scale=0.35]{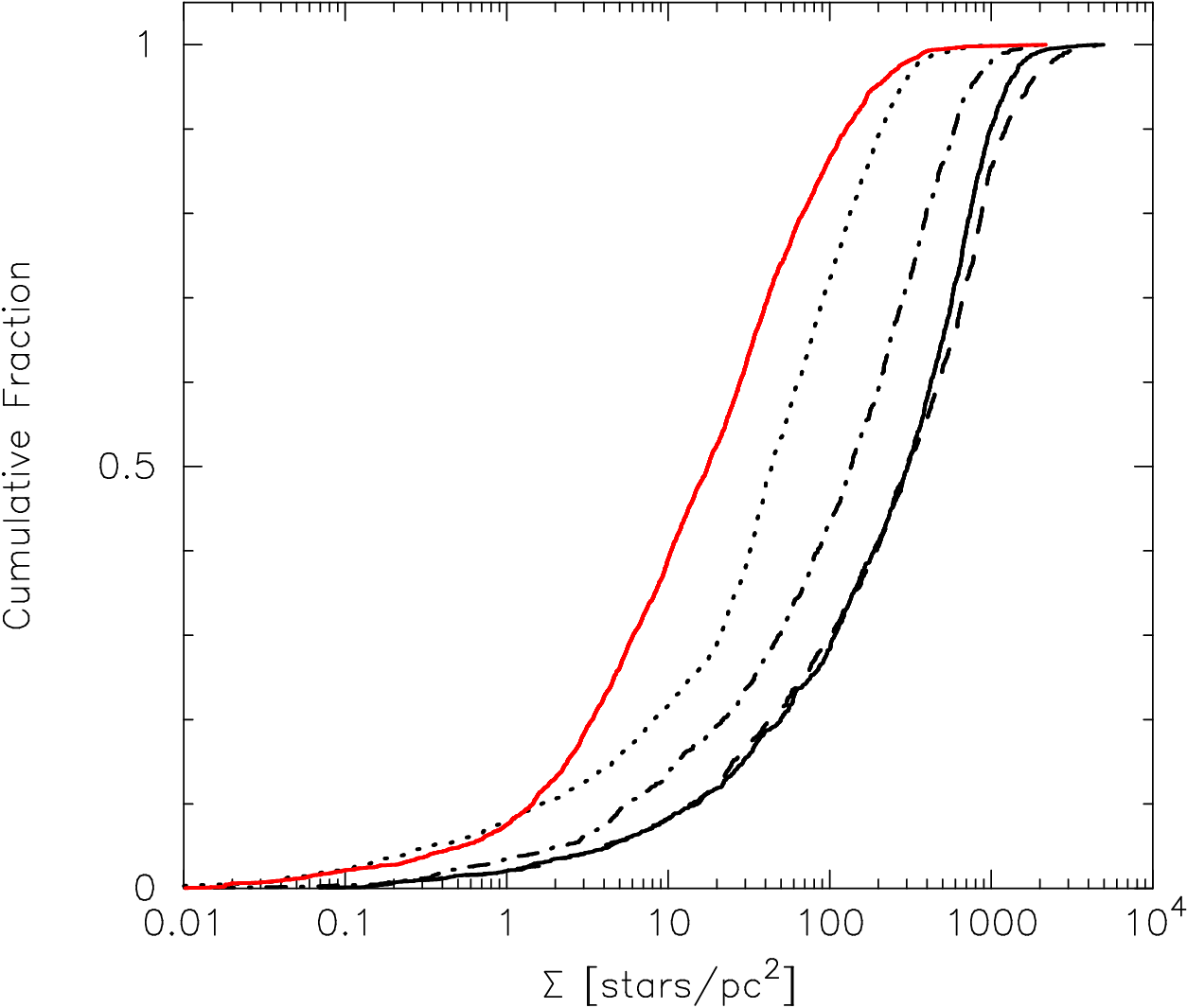}}} 
\hspace*{0.3cm} 
\subfigure[$Q_{\rm vir} = 1.5$ (expanding)]{\label{plummer_evolve-b}\rotatebox{270}{\includegraphics[scale=0.35]{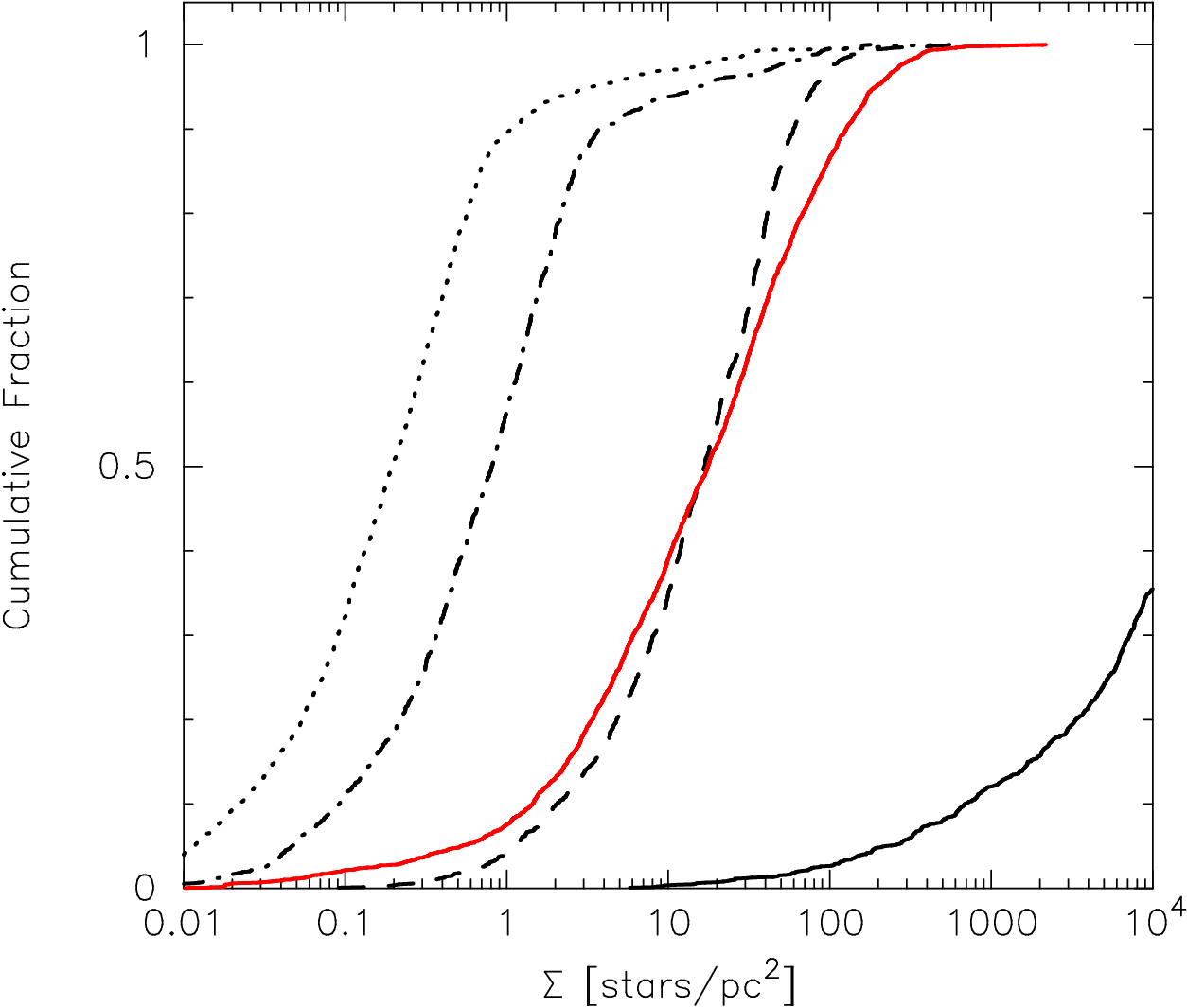}}} 
\caption[bf]{The evolution of $\Sigma$ for Plummer spheres. In (a), we show an $N = 1000$ Plummer sphere initially in virial equilibrium ($Q_{\rm vir}$ with a half-mass radius of 0.8\,pc.  In (b) we show the evolution 
of a `Hot' Plummer sphere, half-mass radius = 0.1pc, virial ratio $Q_{\rm vir} = 1.5$. The data from Bressert et al are shown in red, and the distribution of $\Sigma$ for the clusters at 0\,Myr, 1\,Myr, 5\,Myr 
and 10\,Myr are shown by the solid, dashed, dash-dotted and dotted lines, respectively.}
\label{plummer_evolve}
  \end{center}
\end{figure*}

In this section we show the cumulative distribution of $\Sigma$ for clusters that undergo dynamical evolution. We calculate the cumulative distribution of $\Sigma$ before the cluster evolves (i.e.\,\,at 0\,Myr), 
then at 1\,Myr, 5\,Myr and 10\,Myr. We first examine the evolution of smooth, spherical Plummer spheres in two dynamical states; virial equilibrium ($Q_{\rm vir} = 0.5$) and supervirial, or `warm' ($Q_{\rm vir} = 1.5$). We then 
apply the same analysis to a highly substructured fractal, both subvirial (undergoing cool collapse with $Q_{\rm vir} = 0.3$) and supervirial `warm' ($Q_{\rm vir} = 1.5$). 

Additionally, we calculate the $\mathcal{Q}$-parameter and the evolution of the virial ratio 
as a function of time for the clusters that are initially substructured. When analysing dynamical simulations, the $\mathcal{Q}$-parameter can be artificially 
high due to extreme outliers (R.J.~Allison, priv.~comm.), so we remove the outer 2\,per cent of stars in the $N$-body snapshot from its determination.


\subsubsection{Plummer spheres}

In Fig.~\ref{plummer_evolve-a} we show the cumulative distribution of $\Sigma$ for a Plummer sphere cluster in virial equilibrium, where the virial ratio $Q_{\rm vir} = 0.5$. The half-mass radius of this cluster is initially 
$r_{1/2} = 0.8$\,pc, and the $\Sigma$ distributions at 0\,Myr, 1\,Myr, 5\,Myr and 10\,Myr are shown by the solid, dashed, dot-dashed and dotted lines, respectively. The data from \citet{Bressert10} 
are shown by the solid red line in this plot. Although the cluster expands over its lifetime, and the $\Sigma$ distribution moves to lower values, it does not change in the first 1\,Myr. 

Alternatively, star clusters could undergo a period of rapid expansion due to (almost) instantaneous gas expulsion by the first supernovae in the cluster (\citealp[e.g.][]{Bastian06,Goodwin06}, \citealp[though see][]{Kruijssen12}). For 
example, the velocity dispersion of the Orion Nebula Cluster is estimated to be 4.3\,kms$^{-1}$ \citep{Olczak08}, although \citet{Furesz08} find a lower value of 3.1\,kms$^{-1}$, which is higher than if the cluster were in virial equilibrium and suggests that the cluster is expanding 
\citep[see also][]{Gieles11}. We 
model this by setting the virial ratio to be `supervirial' at the start of the simulation. We show the evolution of an initially very dense ($r_{1/2} = 0.1$\,pc), supervirial ($Q_{\rm vir} = 1.5$) Plummer sphere 
in Fig.~\ref{plummer_evolve-b}.

In this second scenario, the $\Sigma$ distribution for the cluster before dynamical evolution lies far to the right of the observed distribution (the solid line in Fig.~\ref{plummer_evolve-b}); however, the cluster rapidly expands, and after 1\,Myr the 
cluster distribution sits on the observed distribution (the dashed line). Following further evolution and expansion of the cluster, the $\Sigma$ distribution lies to the left of the observed distribution after 5\,Myr (the dashed-dotted line) and 10\,Myr 
(the dotted line). 

We note that in both these evolutionary scenarios the clusters retain the same Plummer sphere morphologies (see e.g.~Fig.~\ref{Plummer-a}) throughout their dynamical evolution, 
 and their virial ratios are essentially constant.

\subsubsection{Fractal clusters}

\begin{figure*}
  \begin{center}
\setlength{\subfigcapskip}{10pt}
\hspace*{-1.5cm}\subfigure[Cool collapse; 0\,Myr]{\label{cool_fractal_evolve-a}\rotatebox{270}{\includegraphics[scale=0.34]{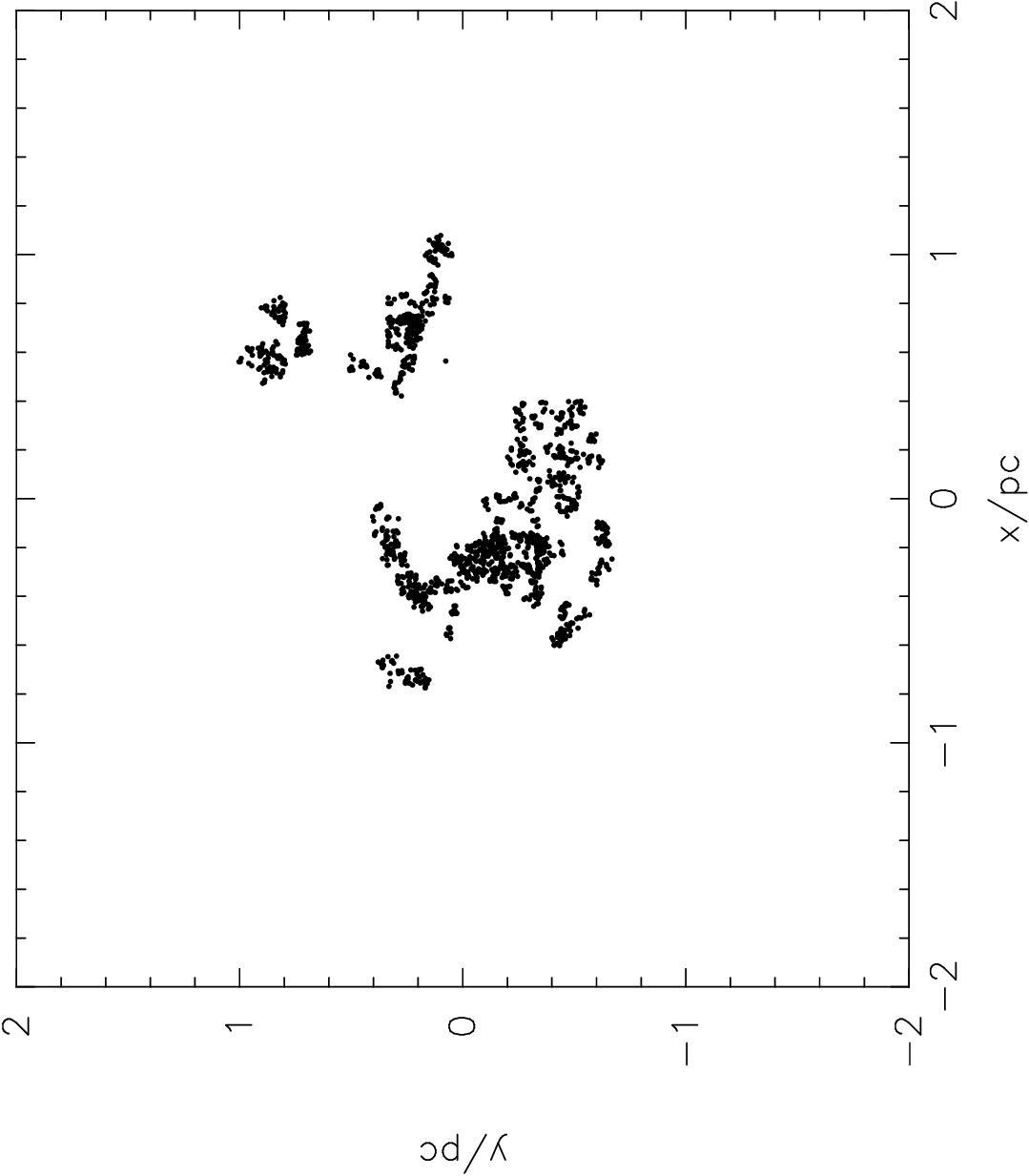}}} 
\hspace*{0.3cm} 
\subfigure[Cool collapse; 5\,Myr]{\label{cool_fractal_evolve-b}\rotatebox{270}{\includegraphics[scale=0.34]{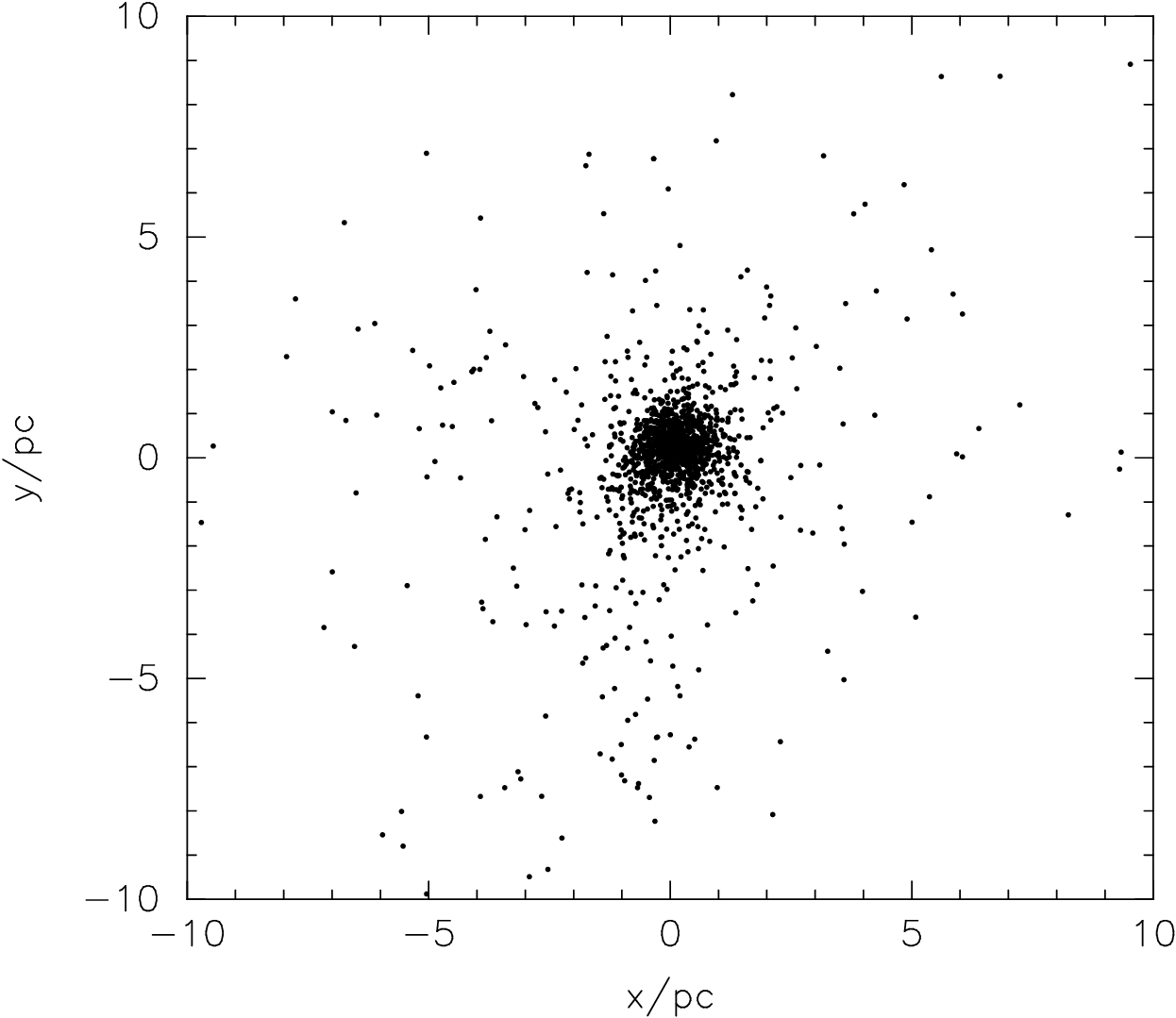}}} 
\hspace*{-1.5cm}\subfigure[Cool collapse; evolution of $\Sigma$]{\label{cool_fractal_evolve-c}\rotatebox{270}{\includegraphics[scale=0.34]{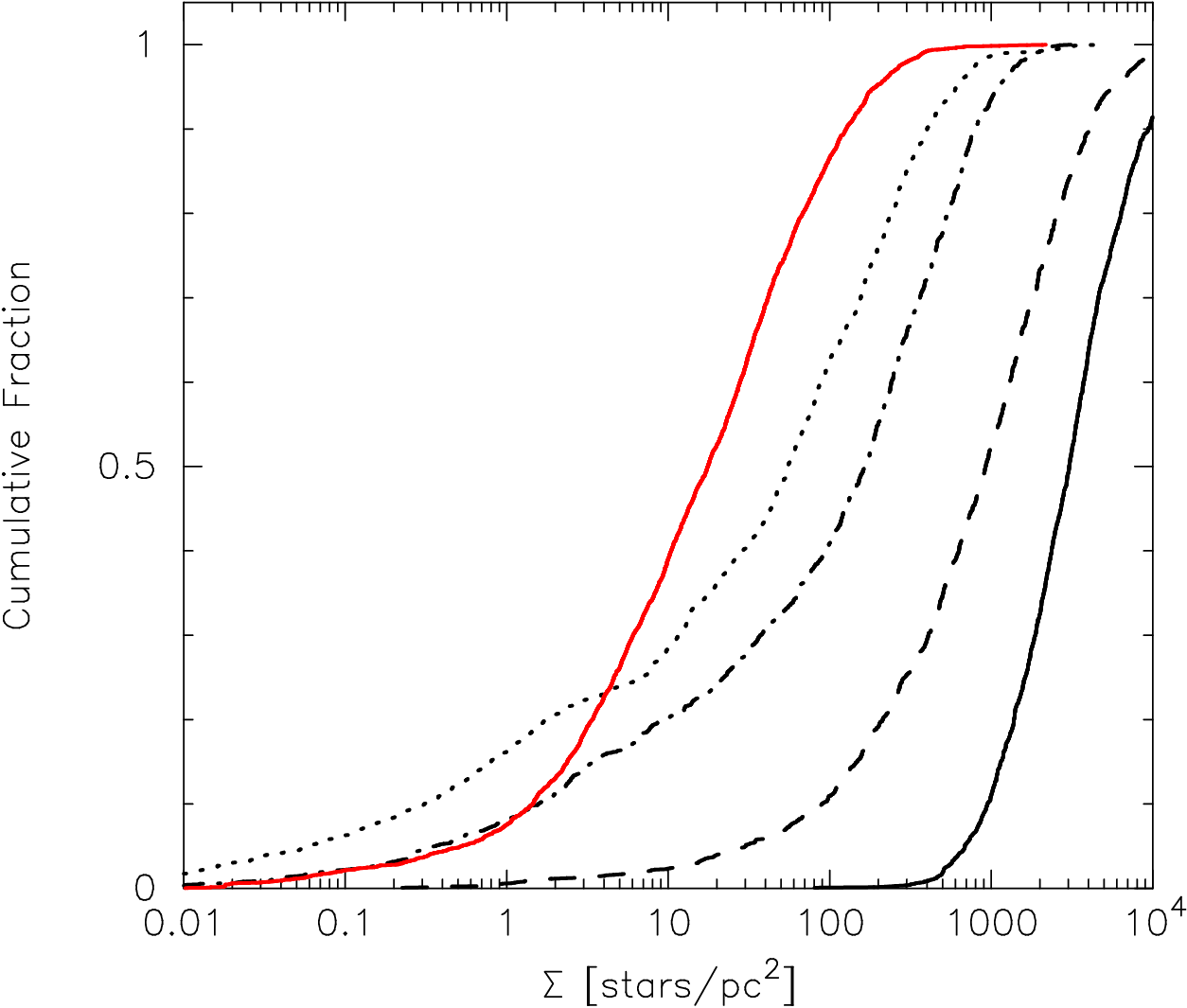}}} 
\hspace*{0.3cm} 
\subfigure[Cool collapse; variation of $\Sigma$ at 5\,Myr]{\label{cool_fractal_evolve-d}\rotatebox{270}{\includegraphics[scale=0.34]{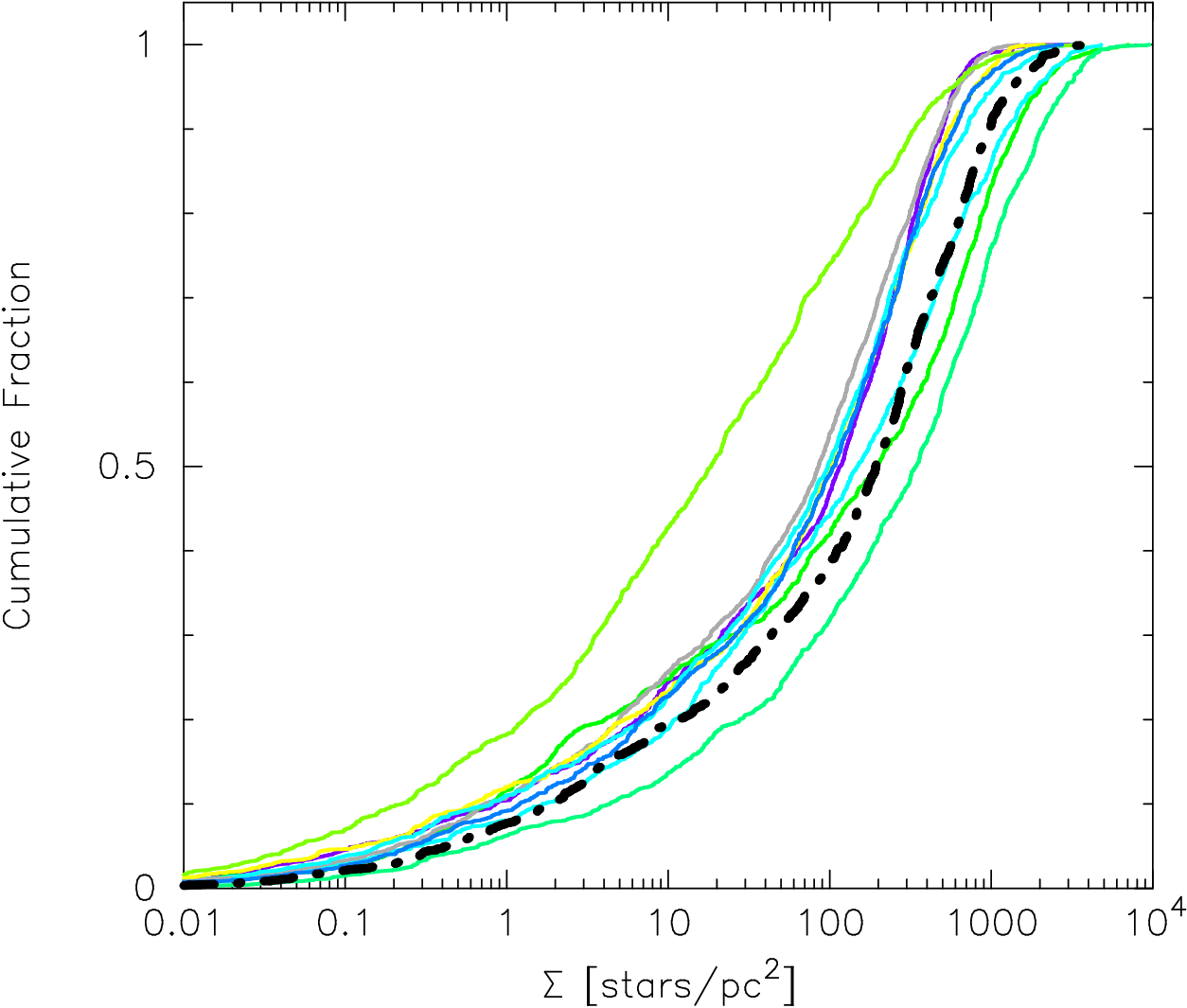}}} 
\hspace*{-1.5cm}\subfigure[Cool collapse; $\mathcal{Q}$--parameter]{\label{cool_fractal_evolve-e}\rotatebox{270}{\includegraphics[scale=0.34]{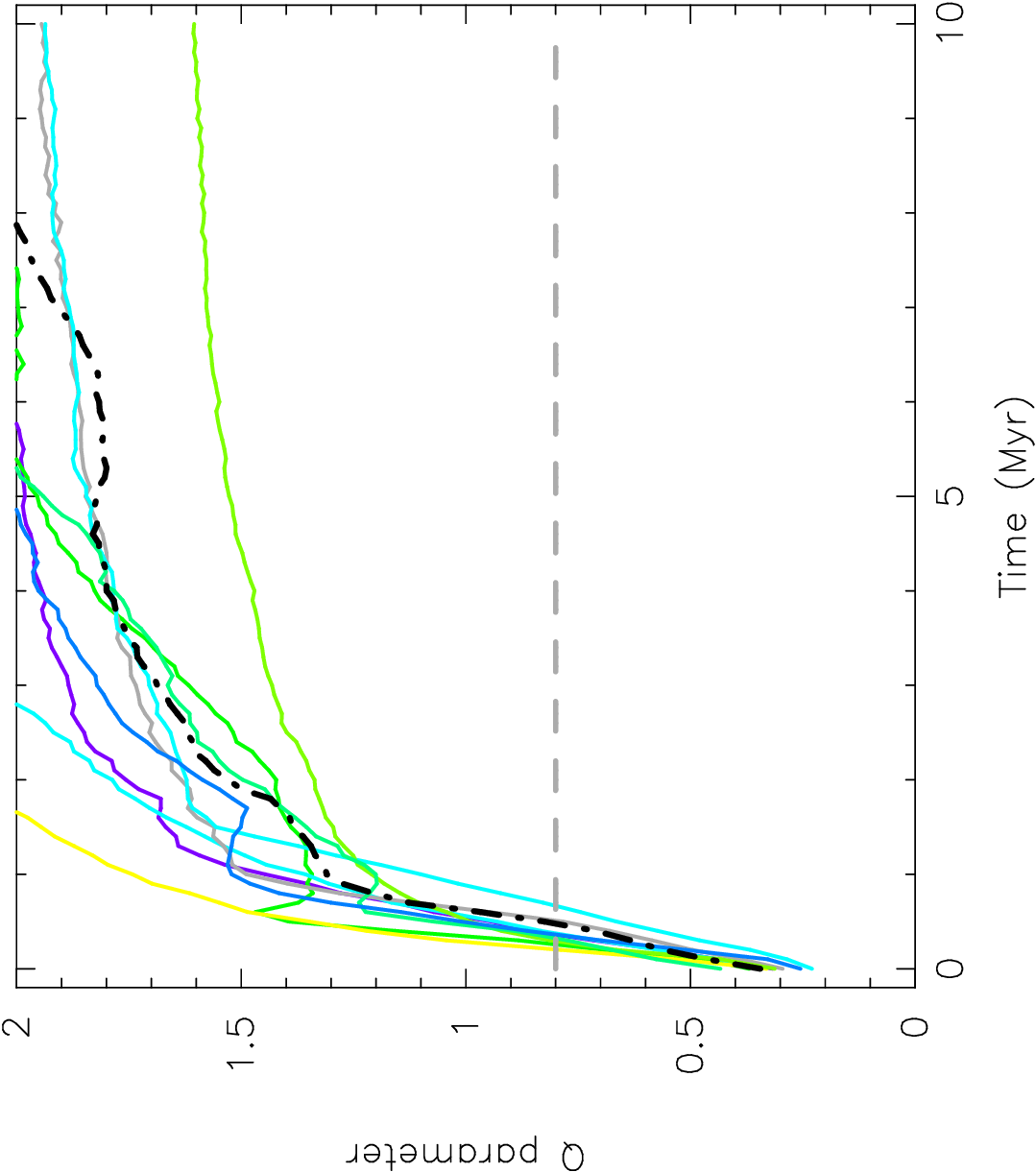}}} 
\hspace*{0.3cm} 
\subfigure[Cool collapse; evolution of virial ratio]{\label{cool_fractal_evolve-f}\rotatebox{270}{\includegraphics[scale=0.34]{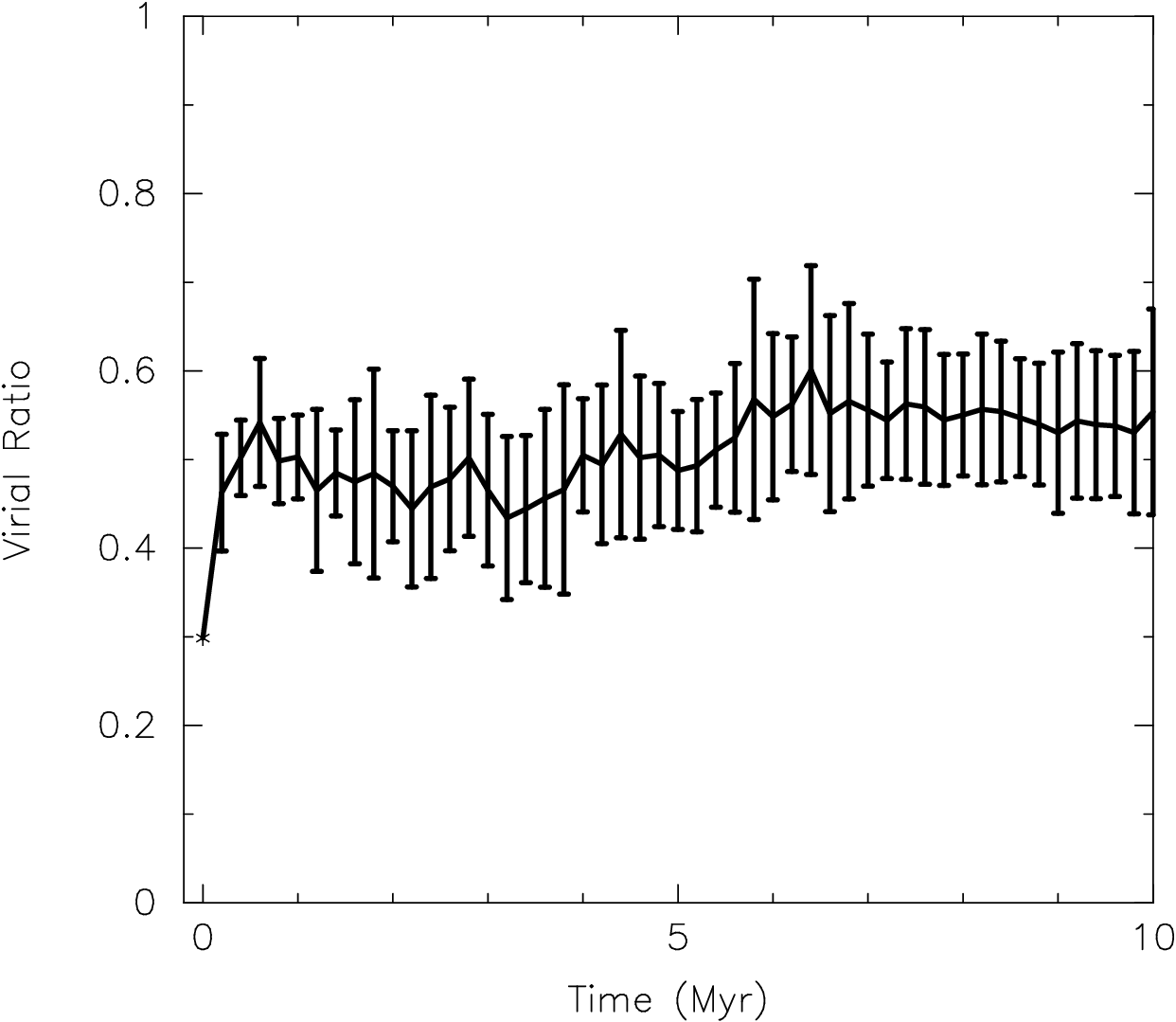}}} 
\caption[bf]{The typical dynamical evolution of a fractal ($D = 1.6$) cluster undergoing cool collapse (virial ratio $Q_{\rm vir} = 0.3$). In panel (a) we show the morphology before evolution, and in panel (b) we show the 
morphology after 5\,Myr. In panel (c) we show the $\Sigma$ distribution at  0\,Myr, 1\,Myr, 5\,Myr and 10\,Myr by the solid, dashed, dash-dotted and dotted lines, respectively and 
the data from Bressert et al are shown by the red line. In panel (d) we show the $\Sigma$ distribution from the simulation in panel (c) at 5\,Myr (the dot-dashed line) compared to the other 9 simulations at 5\,Myr. In panel (e) we show the evolution of the $\mathcal{Q}$--parameter \citep{Cartwright04} during the first 10\,Myr of this cluster's evolution (the dot-dashed line, and all other simulations are also shown for comparison). Finally, we show the average evolution of the virial ratio in the simulations (with 1--$\sigma$ uncertainties) in panel (f).}
\label{cool_fractal_evolve}
  \end{center}
\end{figure*}

Recently, it has been postulated that the Orion Nebula Cluster (ONC) may have been initially subvirial, and undergone `cool' collapse, leading to dynamical mass segregation within 1\,Myr and the formation of Trapezium-like systems \citep{Allison09b,Allison10,Allison11}. 
Such a scenario is also consistent with the observed binary separation distribution \citep{Parker11c}. This scenario favours the formation of stars in the cluster with a high degree of substructure, as a radially smooth cluster cannot mass segregate within the age of 
the ONC, even with subvirial initial conditions \citep{Bonnell98}. 

We show the evolution of the $\Sigma$ distribution for a highly substructured cluster undergoing cool collapse in Fig.~\ref{cool_fractal_evolve}. The cluster has a radius of 1\,pc, fractal dimension $D = 1.6$ and a `cool' virial ratio ($Q_{\rm vir} = 0.3$). In panels (a) and (b) 
we show the morphology of a typical cluster at (a) 0\,Myr and after 5\,Myr of dynamical evolution (b). In a cluster undergoing this dynamical scenario, the stars in the clumpy areas of the fractal interact, and the cluster collapses to a central concentration after 1\,Myr. 
This central concentration has a similar appearance to a Plummer sphere or King profile \citep{Allison10}, and the cluster then relaxes and expands over the following 10\,Myr. Note that the cluster obtains a maximum \emph{central} density after $\sim$1\,Myr \citep{Parker11c}, but Fig.~\ref{cool_fractal_evolve-c} shows that the distribution 
of $\Sigma$ has already moved to lower densities compared to the distribution at 0\,Myr. This is due to ejections from the substructure before the collapse of the cluster, which shifts the $\Sigma$ distribution to lower densities.

In Fig.~\ref{cool_fractal_evolve-c} the initial $\Sigma$ distribution is shown by the solid line. After 1\,Myr the cluster has already reached its densest phase and is now expanding, and the distribution of $\Sigma$ moves to the left (the dashed line). As the cluster 
dissolves the $\Sigma$ distribution widens, and we show the distributions after 5\,Myr (the dot-dashed line) and 10\,Myr (the dotted line). The data for all clusters in the Solar Neighbourhood \citep{Bressert10} is shown by the solid red line. Fractal clusters evolve very stochastically \citep{Allison10,Parker12b} and there is a variation of $\Sigma$ due 
to this. We show the spread in $\Sigma$ at 5\,Myr in our 10 simulations in Fig.~\ref{cool_fractal_evolve-d}. We show the particular cluster in 
Figs.~\ref{cool_fractal_evolve-a}--\ref{cool_fractal_evolve-c} by the thick dot-dashed line, and the remaining simulations by the coloured lines.

We also show the evolution of the $\mathcal{Q}$--parameter in Fig.~\ref{cool_fractal_evolve-e}. Again, the cluster in Figs.~\ref{cool_fractal_evolve-a}--\ref{cool_fractal_evolve-c}  is shown by the thick dot-dashed line. We choose this cluster as a `typical' example because neither the $\Sigma$ 
distribution nor the evolution of the $\mathcal{Q}$--parameter is an outlier, though one could also choose other examples which fit this criteria. The initially substructured cluster has $\mathcal{Q}<0.5$, and the dynamical evolution in our simulation quickly erases this substructure within the first 0.5\,Myr before reaching a large ($>1$) $\mathcal{Q}$--parameter, indicating a centrally concentrated geometry. 

\begin{figure*}
  \begin{center}
\setlength{\subfigcapskip}{10pt}
\hspace*{-1.5cm}\subfigure[Warm expansion; 0\,Myr]{\label{w_fractal_evolve-a}\rotatebox{270}{\includegraphics[scale=0.34]{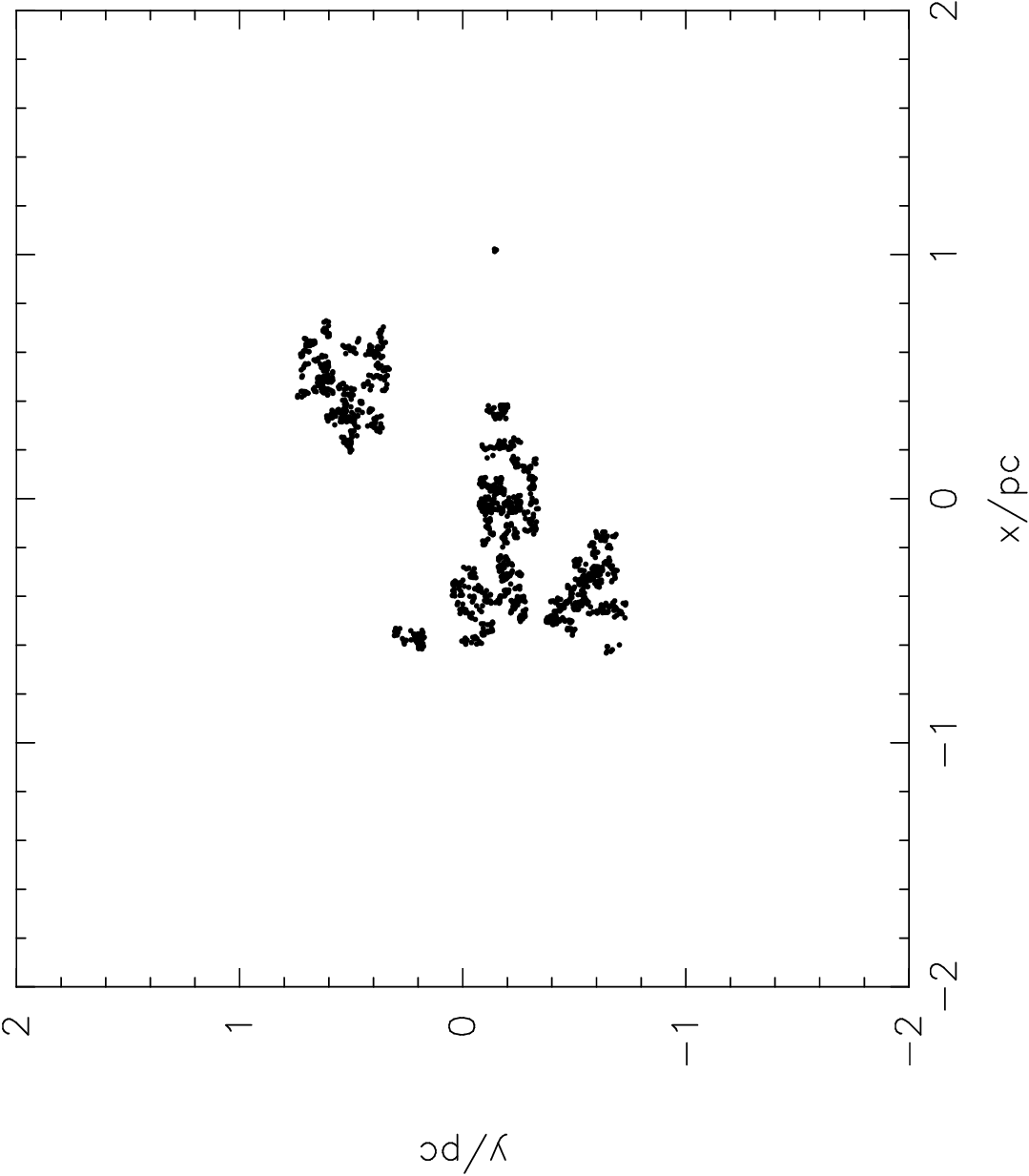}}} 
\hspace*{0.3cm} 
\subfigure[Warm expansion; 5\,Myr]{\label{w_fractal_evolve-b}\rotatebox{270}{\includegraphics[scale=0.34]{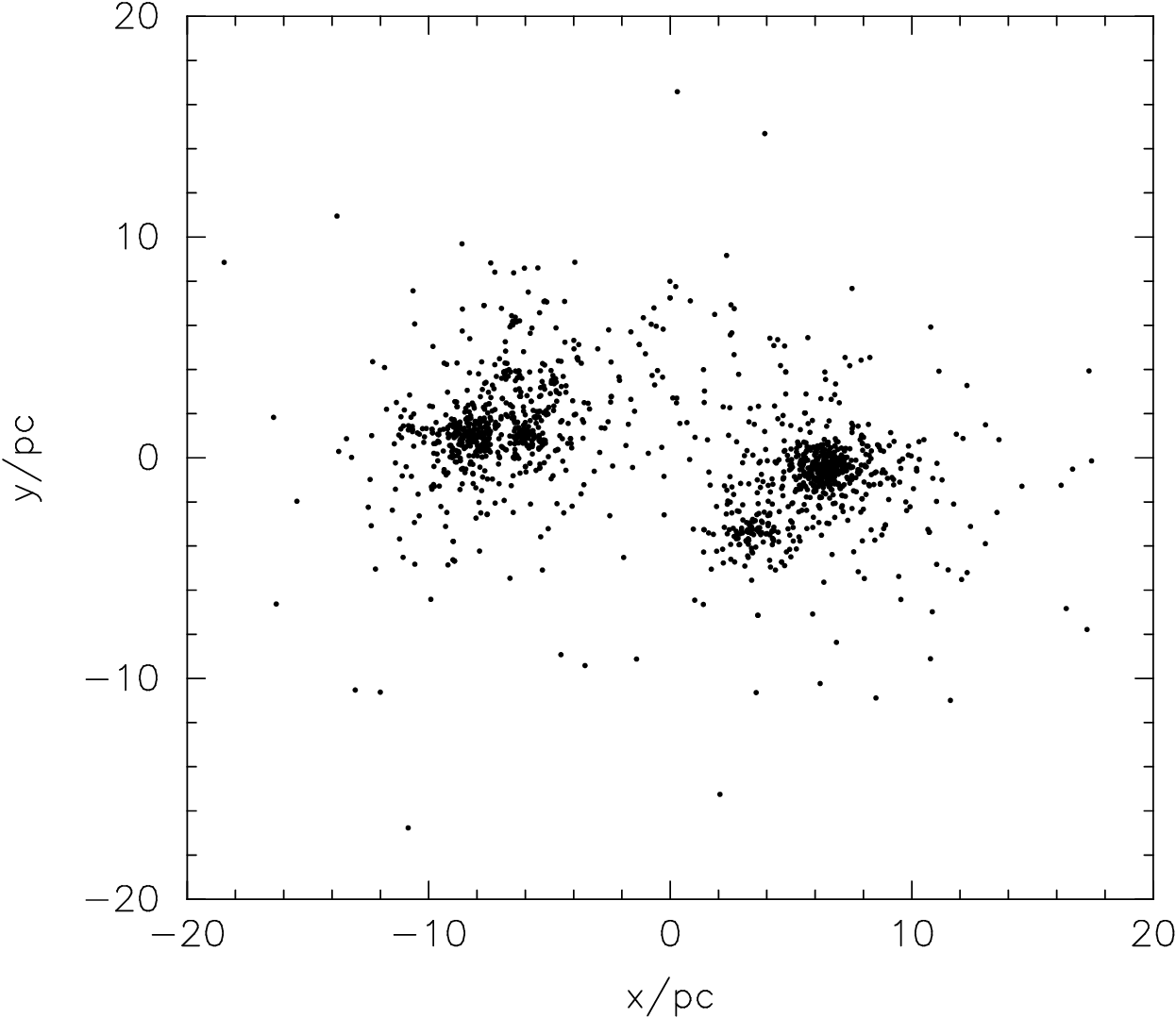}}} 
\hspace*{-1.5cm}\subfigure[Warm expansion; evolution of $\Sigma$]{\label{w_fractal_evolve-c}\rotatebox{270}{\includegraphics[scale=0.34]{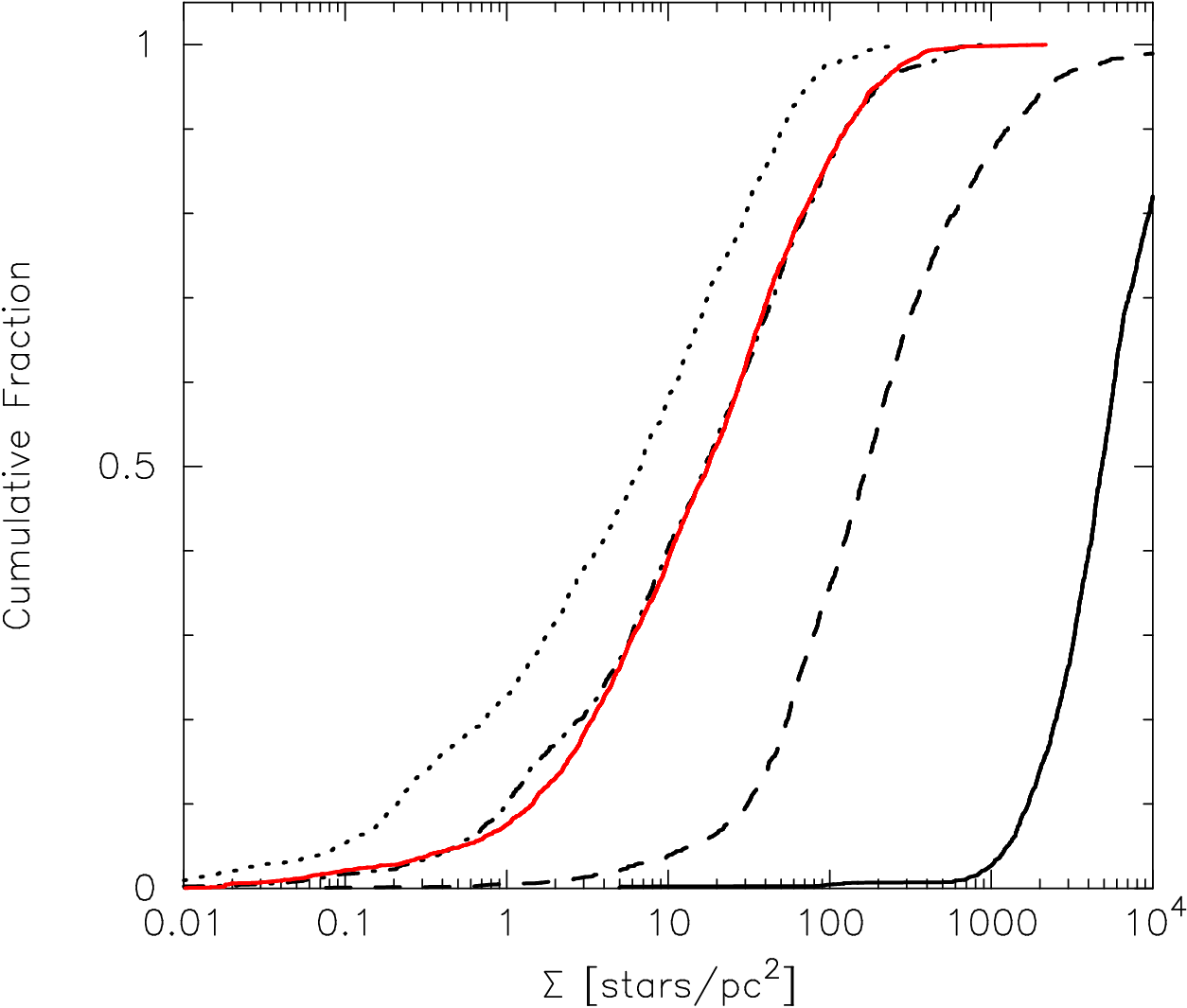}}} 
\hspace*{0.3cm} 
\subfigure[Warm expansion; variation of $\Sigma$ at 5\,Myr]{\label{w_fractal_evolve-d}\rotatebox{270}{\includegraphics[scale=0.34]{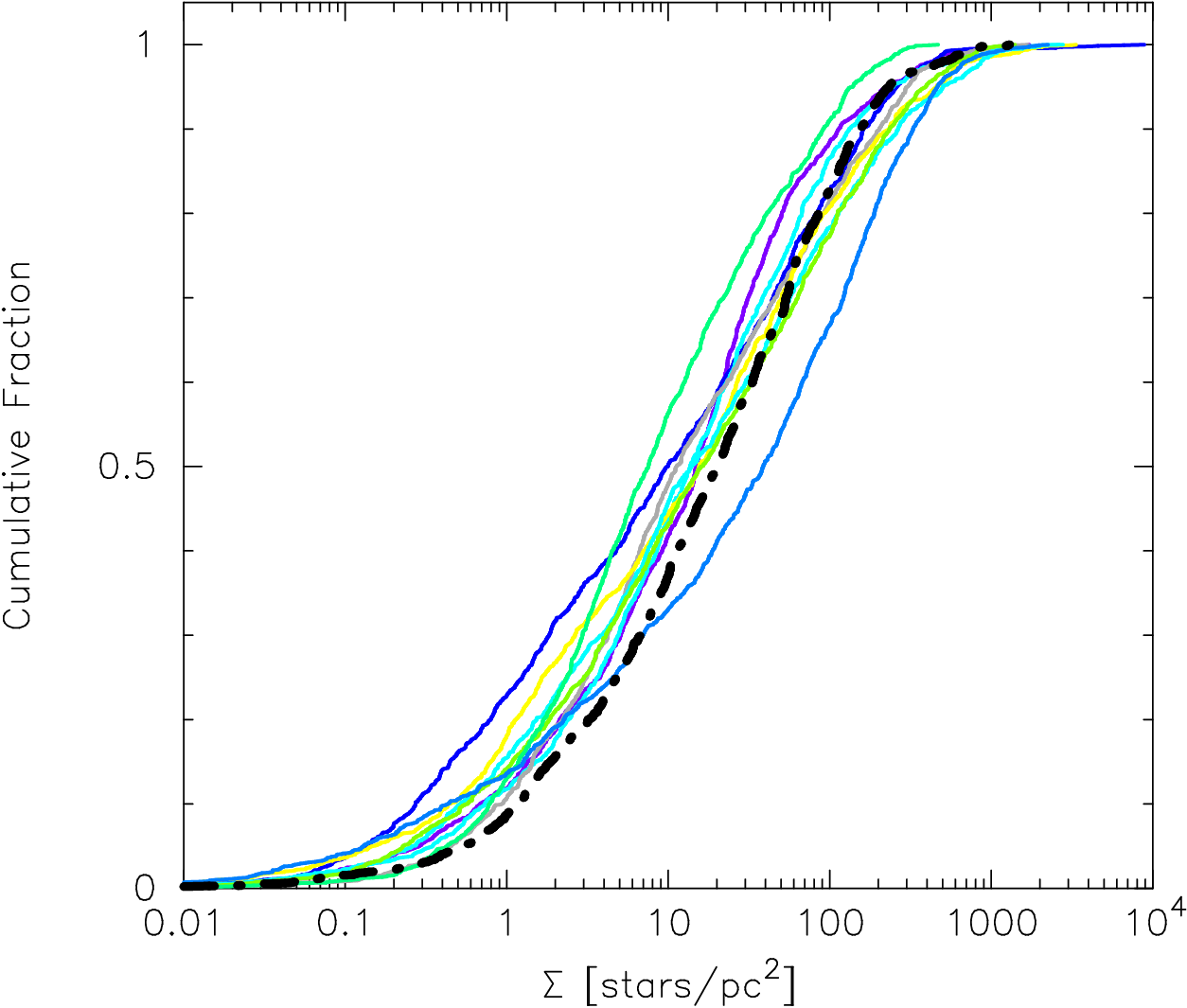}}} 
\hspace*{-1.5cm}\subfigure[Warm expansion; $\mathcal{Q}$--parameter]{\label{w_fractal_evolve-e}\rotatebox{270}{\includegraphics[scale=0.34]{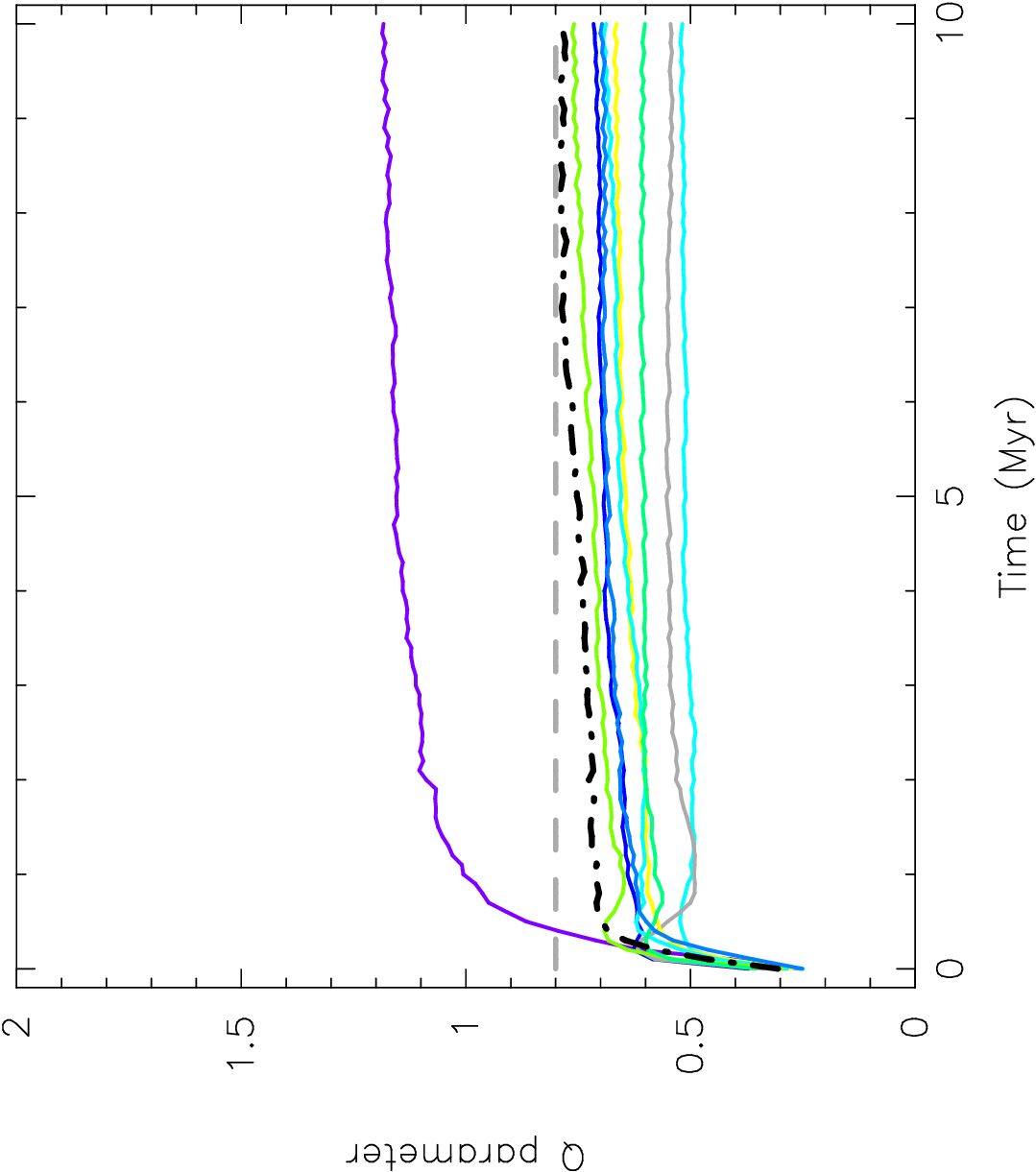}}} 
\hspace*{0.3cm} 
\subfigure[Warm expansion; evolution of virial ratio]{\label{w_fractal_evolve-f}\rotatebox{270}{\includegraphics[scale=0.34]{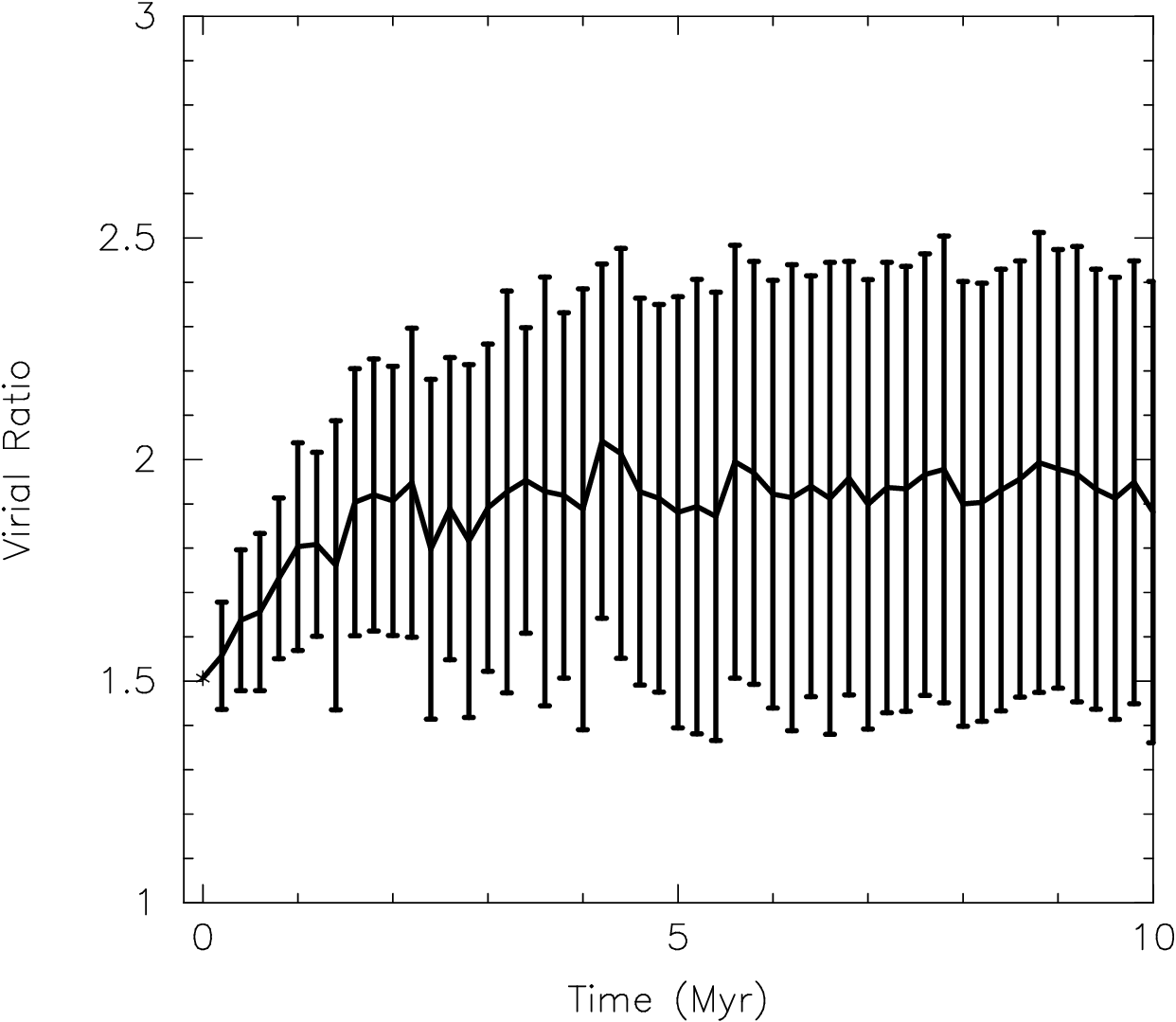}}} 
\caption[bf]{The typical dynamical evolution of a fractal (D = 1.6) cluster undergoing warm expansion (virial ratio $Q_{\rm vir} = 1.5$). In panel (a) we show the morphology before evolution, and in panel (b) we show the 
morphology after 5\,Myr. In panel (c) we show the $\Sigma$ distribution at  0\,Myr, 1\,Myr, 5\,Myr and 10\,Myr by the solid, dashed, dash-dotted and dotted lines, respectively and 
the data from Bressert et al are shown by the red line. In panel (d) we show the $\Sigma$ distribution from the simulation in panel (c) at 5\,Myr (dot-dashed line) compared to the other 9 simulations at 5\,Myr. In panel (e) we show the evolution of the $\mathcal{Q}$--parameter \citep{Cartwright04} during the first 10\,Myr of this cluster's evolution (the dot-dashed line, and all other simulations are also shown for comparison). Finally, we show the average evolution of the virial ratio in the simulations (with 1--$\sigma$ uncertainties) in panel (f).} 
\label{warm_fractal_evolve}
  \end{center} 
\end{figure*}

Finally, we show the evolution of the virial ratio for this collapsing, substructured cluster in Fig.~\ref{cool_fractal_evolve-f}. The cluster begins subvirial ($Q_{\rm vir}$ = 0.3), but rapidly reaches virial equilibrium ($Q_{\rm vir}$ = 0.5) in the first 1\,Myr due to violent relaxation of the substructure, and the subsequent global collapse of the cluster 
\citep[see also][]{Allison10}. The 
behaviour of the virial ratio is qualitatively similar for all 10 simulations (unlike the evolution of the $\mathcal{Q}$--parameter) and we plot the average of all 
10 simulations, with 1--$\sigma$ uncertainties, in Fig.~\ref{cool_fractal_evolve-f}.


The evolution of a substructured, supervirial ($Q_{\rm vir} = 1.5$) cluster undergoing expansion is shown in Fig~\ref{warm_fractal_evolve}. We show the morphology of the (typical) cluster before dynamical evolution in panel~(a), and the corresponding morphology after 5\,Myr of evolution 
in panel~(b). Again, we choose this simulation because it is not an outlier in terms of its global density ($\Sigma$) or its structure (the evolution of the $\mathcal{Q}$--parameter).  The cluster has expanded rapidly, but the clumps within the fractal have also evolved so that the cluster resembles a large association without a central concentration. This is confirmed by the evolution of the $\mathcal{Q}$--parameter (the dot-dashed line Fig.~\ref{w_fractal_evolve-e}), which remains 
below the substructure cut--off value of 0.8 throughout the cluster's evolution. 

We show the evolution of the $\Sigma$ distribution for this cluster in Fig.~\ref{w_fractal_evolve-c}. The distribution quickly widens, and at 5\,Myr sits on the observed distribution (the 
dot-dashed line). In this evolutionary scenario, the median $\Sigma$ value decreases by over three orders of magnitude during the first 10\,Myr of evolution. The spread in 
$\Sigma$ at 5\,Myr between the 10 simulations is shown in  Fig.~\ref{w_fractal_evolve-d}. 

There is also a wide spread in the evolution of the  $\mathcal{Q}$--parameter for these expanding, clumpy clusters. Nine simulations remain substructured 
($\mathcal{Q} < 0.8$) throughout the cluster's lifetime, but in one simulation the substructure evolves to a centrally concentrated profile (the (upper) purple line in 
Fig.~\ref{w_fractal_evolve-e}). There is also some variation in the virial ratio, and we plot the average of all 10 simulations, with 1--$\sigma$ uncertainties, in Fig.~\ref{w_fractal_evolve-f}. All 
the clusters remain supervirial, but some more so than others, hence the large uncertainties.

Finally, we note that for both the cluster undergoing cold collapse, and the rapidly expanding cluster, the $\Sigma$ distribution for the fractal widens. Therefore, if stars were to form in a fractal distribution, it would be difficult to rule out this formation scenario by analysing the shape of the 
$\Sigma$ cumulative distribution.

Note that we would expect a great deal of dynamical interaction in the substructure as the supervirial cluster expands. Even though it does not reach a centrally concentrated dense phase (like the initially subvirial cluster), pockets of substructure are 
sufficiently dense enough to process binaries \citep{Parker11c}, and planetary systems \citep{Parker12a}.

\section{Discussion}
\label{discuss} 

In Section~\ref{results} we presented the cumulative distribution for stellar surface densities, $\Sigma$, for a range of different single cluster morphologies. We divide the morphologies into two subgroups; radially smooth (the Plummer spheres and 
King profiles), and substructured (the fractals and associations). We have deliberately chosen the radii of the clusters so that the median of the $\Sigma$ distribution is overlaid upon the median of the ensemble observational data presented in \citet{Bressert10}. 

From inspection, the radially smooth clusters are able to reproduce the shape of the observed cumulative distribution, and the King profile, which is more centrally concentrated than the Plummer sphere, is an especially good fit. On the other hand, 
the clusters that are initially substructured have a narrower cumulative distribution than the observed data. We note that varying the parameters of these clusters (overall radius, fractal dimension, association node radius) does not widen this 
distribution. 

We have also demonstrated, with the KS test, that only one single cluster morphology, a King profile, is consistent with the dataset of \citet{Bressert10}. This is not surprising, since the observational dataset includes stars from different star forming regions, some of which 
contain substructure, and others have a more centrally concentrated, smooth radial profile. For example, within the dataset are Taurus and Ophiuchus. A study of substructure using the $\mathcal{Q}$--parameter by \citet{Cartwright04} showed Ophiuchus 
to have a smooth radial profile ($\mathcal{Q} = 0.85$), whereas Taurus is substructured, with a low fractal dimension ($\mathcal{Q} = 0.45$). 

\citet{Bressert10} point out that the smooth cumulative distribution in the data is evidence of hierarchical star formation, rather than a bi-modal distribution of clusters and associations. We demonstrate this in Figs.~\ref{bg-a}--\ref{bg-c}, where we place an 
association in a much more sparse background. The histogram displays obvious bimodality, and the cumulative distribution shows a prominent kink, which would be observed if star formation was heavily bi-modal. However, we also caution that if this distinction were to be blurred, then 
a $\Sigma$ distribution would not be able to detect a more diffuse association against a background of field stars, as we demonstrate in Figs.~\ref{bg-d}--\ref{bg-f}.

We have also shown the effects of dynamical evolution of the clusters on the distribution of $\Sigma$. Firstly, we have modelled a Plummer sphere with a similar half-mass radius to that on the ONC today \citep[0.8\,pc,][]{Hillenbrand98}. Even for a cluster 
in virial equilibrium, over a period of 10\,Myr the distribution of $\Sigma$ evolves to lower densities as the cluster expands through interactions. In Fig.~\ref{plummer_evolve-a} we show the initial distribution (the solid black line), and the 
distributions at 1, 5, and 10\,Myr (the dashed, dot-dashed and dotted lines, respectively). The distributions at 0~and~1\,Myr are virtually identical, and we note that the distribution of $\Sigma$ observed by \citet{Bressert10} 
is for young ($\leq$1\,Myr) star forming regions. Therefore, if all star formation was to occur in virial equilibrium then we would expect the distribution of $\Sigma$ to be indicative of the primordial density of the star forming region for 
which it was measured.

Secondly, we have modelled an initially very dense Plummer sphere with a half-mass radius of only 0.1\,pc, but with an initially supervirial ratio ($Q_{\rm vir} = 1.5$), so that the cluster immediately expands. It has been argued that rapid gas expulsion caused by the first 
supernovae or the collective effect of winds from OB stars in a star cluster can cause the expansion  of a cluster \citep[e.g.][and references therein]{Tutukov78,Bastian06,Goodwin06,Bastian08}, although this effect may not be as significant as first thought \citep[e.g.][]{Bastian11,Kruijssen12}. We show the effect of this scenario on the $\Sigma$ distribution as a function of time in Fig.~\ref{plummer_evolve-b}. 
The cluster is initially very dense, but rapidly expands so that after 1\,Myr the density is similar to that observed for local star forming regions by \citet{Bressert10}.

Recently, a dynamical evolution scenario for the ONC has been developed, in which the cluster is initially substructured, and subvirial. The cluster collapses to a centrally  concentrated sphere which appears very similar to a King or Plummer profile. This scenario can 
allow dynamical mass segregation of the highest mass stars within the first 1\,Myr \citep{Allison09b,Allison10}, and the formation of Trapezium-like systems \citep{Allison11}. Furthermore, such initial conditions heavily process a primordial binary population 
\citep[which is consistent with the observed separation distribution and binary fraction in the ONC,][]{Parker11c} and can affect planetary systems \citep{Parker12a}. The distribution of $\Sigma$ for such a cluster is shown in Fig.~\ref{cool_fractal_evolve}. The cluster rapidly 
collapses, and begins expanding after 1\,Myr. The increased tail in the distribution at low densities is due to ejections from within the substructure. Whilst 
the change in the $\Sigma$ distribution with time is not as extreme as in the supervirial Plummer sphere, it still demonstrates that if a cluster were to evolve in this way, the measured $\Sigma$ value at 1\,Myr is highly unlikely to be primordial or indicative of the amount of previous dynamical processing.

For completeness, we have also shown the evolution of the $\Sigma$ distribution in an initially supervirial fractal (Fig.\ref{warm_fractal_evolve}). The overall evolution is similar to the supervirial Plummer sphere, though not as extreme, due to the much lower initial density of the 
fractal resulting in fewer violent ejections. Unlike the subvirial fractal, this cluster retains its substructure as it evolves to form an association, and the structural difference between these clusters is highlighted by the evolution of the $\mathcal{Q}$--parameter in Figs.~\ref{cool_fractal_evolve-d}~and~\ref{w_fractal_evolve-d}.

The measurement of the $\Sigma$ distribution for various dynamical scenarios has shown that, at  best, it reflects the state of the cluster or star forming region now, and in the absence of further information should not be used to make inferences on the birth environment 
of stars or planets. Figures~\ref{plummer_evolve},~\ref{cool_fractal_evolve}~and~\ref{warm_fractal_evolve} demonstrate that the distribution is likely to drastically change in the first 5~Myr or so of the cluster's lifetime, unless all stars form in a relatively sparse cluster in virial equilibrium  (recent work by \citet*{Gieles12} and \citet{Moeckel12} has also 
pointed out that the $\Sigma$ distribution evolves to lower densities through dynamical interactions).  

However, the $\Sigma$ distribution used in tandem with the $\mathcal{Q}$-parameter and age estimates \emph{may} provide information on the initial density and level of substructure. As substructure is erased quickly through dynamics \citep{Allison10}, if the $\mathcal{Q}$-parameter is high ($\mathcal{Q} > 0.8$) we can 
postulate that dynamical evolution may have occurred and the present density may not be primordial. Conversely, as young, unevolved regions are likely to be substructured \citep{Sanchez09}, then if the $\mathcal{Q}$-parameter is low ($\mathcal{Q} < 0.8$) the the $\Sigma$ distribution likely reflects 
the primordial density of that cluster or star forming region. 


The volume-limited census of star forming regions in the local Solar neighbourhood presented by \citet{Bressert10} suggests that most regions would not be dense enough to affect the formation and evolution of binary stars and planetary systems (especially if these regions formed in virial equilibrium). Unfortunately, no data from the 
central region of the ONC is included 
in the sample of stars from Orion in \citet{Bressert10} due to crowding of stars in the images. The contribution of the Orion region in their sample is 2696 out of a total of 3857 stars. If the central region of the ONC was included, then the number of stars from Orion would increase by $\sim$1000 and it is unclear how this would change 
the $\Sigma$ distribution in the local neighbourhood \citep[though it would likely contribute to the high-density region of this distribution,][]{Bate98b}. The arguments for a dynamically active ONC, presented by \citet{Kroupa99,Scally05,Parker09a,Allison09b,Allison10,Allison11,Parker11c}, suggest that this cluster underwent dynamical interactions that have affected binary stars and planetary systems.

If one assumes the uniform 
cluster mass function of $N(M) \propto M_{\rm clus}^{-2}$ \citep{Lada03}, and also that clusters are uniformly disrupted (independent of mass), then the same mass of stars enters the Galactic field from one very high mass cluster, as does stellar mass from many smaller clusters with a combined stellar 
mass equal to the high-mass cluster. 


We can postulate that the local Galactic field may be populated by the sum of ONC-like,  and low-mass star-forming regions (if all of these clusters eventually disperse). Thus, a different interpretation of the $\Sigma$ distribution is that the dominant star forming event 
in terms of mass is the ONC, which contributes 25\,per cent of mass to the field. If the ONC is dynamically old \citep{Kroupa99,Allison10,Parker11c} then the stars from the ONC with low surface densities \emph{today} may have been in a much  more dense region in the past (recall e.g.~Fig. \ref{cool_fractal_evolve}, 
where a fractal collapses to a central concentration, ejecting a halo of stars in the process \citep{Allison12}).

We therefore suggest that as a lower limit,  $\sim$25\,per cent of stars originate in a dense cluster, but primordial substructure in star-forming regions can potentially disrupt binaries and planets without the need for high global densities \citep{Parker11c,Parker12a}. This substructure 
could increase this fraction of dynamically susceptible systems to $\sim$50\,per cent. Such a scenario would then have implications for the formation and survival of binaries \citep[e.g.][]{Kroupa99,Marks11,Parker11c} and planetary systems \citep{Bonnell01b,Scally01,Adams06,Olczak08,Parker09c,Parker12a}.

Finally, we suggest that in order to determine both the current and initial density, and structure of a star forming region, a combination of tools must be used. A low, global 
density does not necessarily preclude the possibility of locally dense substructure, which can be identified using the $\mathcal{Q}$--parameter. Similarly, a low global density 
now does not rule out the possibility than the cluster was originally several orders of magnitude more dense at an earlier time. In the near future, accurate measurements of the positions and velocities of stars using data from the GAIA mission and its associated spectroscopic surveys \citep[e.g.][]{Randich12}, in tandem with numerical simulations, will enable a detailed dynamical history of nearby star forming regions to be made \citep[e.g.][]{Allison12}. 

\section{Conclusions}
\label{conclude}

In this paper, we have modelled both static and evolving star clusters with different morphologies, and determined the distribution of stellar surface densities, $\Sigma$, and the $\mathcal{Q}$-parameter. In our $N$-body simulations 
we determine the $\Sigma$ distribution at 0, 1, 5 and 10\,Myr. Our conclusions are the following:

(i) The distribution of $\Sigma$ is degenerate. Many different morphologies reproduce a smooth, continuous distribution of stellar densities. We find that a substructured star forming region (either a fractal, or an association) 
has a narrower cumulative distribution than a centrally concentrated spherical cluster, such as a Plummer sphere or a King profile. However, these morphologies can be much more easily distinguished by other methods, 
such as the $\mathcal{Q}$--parameter \citep{Cartwright04}. Characterizing substructure in clusters is important, as this has recently been shown to facilitate dynamical mass segregation and disrupt binary and planetary systems in young clusters \citep{Allison10,Parker11c,Parker12a}.

(ii) No single morphology is a good match to the observed $\Sigma$ distribution presented in \citet{Bressert10}. We interpret this as being due to the superposition of $\Sigma$ values from the different star forming regions 
in the observational sample. 

(iii) The $\Sigma$ distribution can be significantly shifted to lower densities for both substructured and smooth clusters due to early dynamical evolution of the cluster. In the case of a supervirial (expanding) cluster, the median $\Sigma$ value can be similar to 
the observed distribution after just 1\,Myr of dynamical evolution. The dynamical evolution of a fractal causes the distribution of $\Sigma$ to widen, as well as shifting it to lower densities. Even an initially dense cluster in virial equilibrium will expand, 
due to two-body relaxation and push the $\Sigma$ distribution to lower values in the first 10\,Myr of the cluster's life. 

(iv) Substructure is quickly erased through dynamical interactions, and therefore the $\mathcal{Q}$--parameter reaches high values rapidly. If all star forming regions start substructured, then a low $\mathcal{Q}$--parameter could indicate that the 
$\Sigma$ distribution reflects the primordial density of the region. This local substructure could facilitate dynamical processing of systems, without the need for high global densities.

(v) Although the dynamical scenarios that drastically alter the $\Sigma$ distribution (a collapsing fractal cluster, or supervirial, initially dense Plummer sphere), can be viewed as extrema in terms of the formation and evolution of clusters, they represent 
possible initial conditions for the ONC \citep{Kroupa99,Parker09a,Allison09b,Allison10,Parker11c}, which would contribute $\sim$25\,per cent of stars to the local volume-limited $\Sigma$ distribution if it were to be included in the sample. If primordial 
substructure is taken into account, the fraction of stars that could be affected by dynamics in a cluster could be as high as 50\,per cent.

(vi) If star forming regions form in virial equilibrium, then dynamical relaxation cannot significantly alter the $\Sigma$ distribution. In such systems, the $\Sigma$ distribution is a tracer of the initial density of a star forming region.

The combination of points (iii), (iv) and (v) lead us to suggest that the $\Sigma$ distribution in local star forming regions may indicate a relatively quiescent natal environment for star formation, but the possibility of a more dynamically active initial environment 
for up to 50\,per cent of the stars cannot be ruled out. We suggest that the $\Sigma$ distribution in local star forming regions should not be over-interpreted \citep[see also][]{Bate98b,Gieles12}, and that one should use a range of different metrics (e.g.\,\,a minimum spanning tree analysis 
such as the $\mathcal{Q}$--parameter in tandem with age estimates, KS tests and``inverse binary population synthesis'' \citep{Kroupa95a,Kroupa95b}) to determine the past and present state of star forming regions. Detailed kinematic information from the GAIA mission and its associated spectroscopic surveys will enable a detailed dynamical history of field and cluster stars to be 
made, which could in principle be used to infer the density of the birth environment of most stars.

\section*{Acknowledgements}

We thank the anonymous referee for their comments and suggestions, which improved the original manuscript.  We also thank Eli Bressert and Nate Bastian for kindly providing us with the cumulative distribution from \citet{Bressert10}, and for a thorough critique of an earlier draft of this work. Thanks also to 
Richard Allison for useful discussions. The simulations in this work were performed on the \texttt{BRUTUS} computing cluster at ETH Z{\"u}rich.

\bibliographystyle{mn2e}
\bibliography{general_ref}

\label{lastpage}

\end{document}